# Uncertainty and Error Quantification for Data-Driven Reynolds-Averaged Turbulence Modelling with Mean-Variance Estimation Networks

Anthony Man, Mohammad Jadidi, Amir Keshmiri, Hujun Yin, Yasser Mahmoudi[*]

School of Engineering, The University of Manchester, Manchester, M13 9PL, UK

*Corresponding author: yasser.mahmoudilarimi@manchester.ac.uk

## Abstract

Amid growing interest in machine learning, numerous data-driven models have recently been developed for Reynolds-averaged turbulence modelling. However, their results generally show that they do not give accurate predictions for test cases that have different flow phenomena to the training cases. As these models have begun being applied to practical cases typically seen in industry such as in cooling and nuclear, improving or incorporating metrics to measure their reliability has become an important matter. To this end, a novel data-driven approach that uses mean-variance estimation networks (MVENs) is proposed in the present work. MVENs enable efficient computation as a key advantage over other uncertainty quantification (UQ) methods – during model training with maximum likelihood estimation, and UQ with a single forward propagation. Furthermore, the predicted standard deviation is also shown to be an appropriate proxy variable for the error in the mean predictions, thereby providing error quantification (EQ) capabilities. The new tensor-basis neural network with MVEN integration was compared with its popular underlying data-driven model by evaluating them on two test cases: a separated flow and a secondary flow. In both cases, the proposed approach preserved the predictive accuracy of the underlying data-driven model, while efficiently providing reliability metrics in the form of UQ and EQ. For the purposes of turbulence modelling, this work demonstrates that the UQ and EQ mechanisms in MVENs enable risk-informed predictions to be made and therefore can be insightful reliability measures in more complex cases, such as those found in industry.

# Nomenclature

| Latin Letters | Meaning | Unit |
|---|---|---|
| $b_{ij}$, $\boldsymbol{b}$ | Normalised Reynolds stress anisotropy tensor | - |
| $B$ | Batch size | - |
| $E$ | Mean negative log likelihood loss | - |
| $\boldsymbol{g}$ | Scalar coefficients ($\boldsymbol{g} = [g_1, \ldots, g_{10}]^T$) | - |
| $H_h$ | Periodic hill height | m |
| $H_d$ | Square duct height | m |
| $k$ | Turbulent kinetic energy | m²/s² |
| $\mathcal{M}$ | MVEN model | - |
| $n$ | Number of basis tensors | - |
| $N$ | Total dataset entries | - |
| $p$ | Pressure | kg/ms² |
| $R$ | Dimensional mean rotation rate tensor | 1/s |
| $\boldsymbol{R}$ | Non-dimensional mean rotation rate tensor | - |
| $Re$ | Reynolds number | - |
| $RMSE_{all}$ | Root mean squared error (RMSE) with mean taken over all components and all entries $N$ | - |
| $RMSE_{ij}$ | RMSE with mean taken over all entries $N$ for component prediction $i, j$ | - |
| $RMSE^{(q)}$ | RMSE with mean taken over all component predictions for entries $N$ | - |
| $S$ | Dimensional mean strain rate tensor | 1/s |
| $\boldsymbol{S}$ | Non-dimensional mean strain rate tensor | - |
| $t_R$ | Timescale for mean rotation rate tensor | s |
| $t_S$ | Timescale for mean strain rate tensor | s |
| $\boldsymbol{\mathcal{T}}$ | Basis tensors ($\boldsymbol{\mathcal{T}} = [\boldsymbol{T}_1, \ldots, \boldsymbol{T}_{10}]$) | - |
| $u_i$ | Velocity in the i$^{th}$ direction | m/s |
| $U_b$ | Bulk inlet velocity | m/s |
| $x$ | Arbitrary model scalar input | - |
| $x_i$ | Distance in the i$^{th}$ direction | m |
| $\boldsymbol{x}$ | Arbitrary model vector input | - |
| $y$ | Arbitrary model scalar output | - |
| $\boldsymbol{y}$ | Arbitrary model vector output | - |

| Greek Letters | | |
|---|---|---|
| $\alpha$ | Mixing probability | - |
| $\beta$ | Steepness factor | - |
| $\delta_{ij}$ | Kronecker delta | - |
| $\varepsilon$ | Turbulent kinetic energy dissipation rate | m²/s³ |
| $\boldsymbol{\lambda}$ | Input features | - |
| $\boldsymbol{\mu}$ | Mean (scalar or vector) | - |
| $\nu$ | Kinematic viscosity | m²/s |
| $\rho$ | Density | kg/m³ |
| $\boldsymbol{\sigma}$ | Standard deviation (scalar or vector) | - |
| $\sigma_{\boldsymbol{b}}$ | Standard deviation averaged over all anisotropy tensor components | - |
| $\tau_{ij}$ | Reynolds stress tensor ($\tau_{ij} = \overline{u'_i u'_j}$) | m²/s² |

**Superscript**

| | | |
|---|---|---|
| $\overline{\phantom{x}}$ | Reynolds-averaged | |
| $'$ | Fluctuating quantity | |

|   |   |
|---|---|
| $\hat{}$ | Predicted quantity |
| $(q)$ | Entry count |
| $(t)$ | Batch count |
| $targ$ | Target quantity |

**Abbreviation**

|   |   |
|---|---|
| BH | Boussinesq hypothesis |
| BNN | Bayesian neural network |
| CCC | Concordance correlation coefficient |
| DNS | Direct numerical simulation |
| DS | Dataset shift |
| EQ | Error quantification |
| GEVH | General eddy viscosity hypothesis |
| ITD | Infinite training data |
| LES | Large eddy simulation |
| LEVM | Linear eddy viscosity model |
| ML | Machine learning |
| MLE | Maximum likelihood estimation |
| MLP | Multilayer perceptron |
| MNLL | Mean negative log likelihood |
| MSE | Mean squared error |
| MVEN | Mean-variance estimation networks |
| NLEVM | Non-linear eddy viscosity model |
| NN | Neural network |
| NUM | Non-unique mapping |
| PCC | Pearson correlation coefficient |
| PDF | Probability density function |
| RANS | Reynolds-averaged Navier Stokes |
| RMSE | Root mean squared error |
| SST | Shear stress transport |
| TBNN | Tensor basis neural network |
| TKE | Turbulent kinetic energy |
| UQ | Uncertainty quantification |

# 1 Introduction

Modelling of turbulent flows for industrial applications has been dominated by Reynolds-averaged Navier-Stokes (RANS) approaches for decades [1]. Although advancement in computational power over recent years has driven the development of scale-resolved methods such as direct numerical simulation (DNS), they are still too computationally expensive for many practical cases [2]. Therefore, RANS models remain commonly used for industrial simulations. Among the different classes of RANS models, linear eddy viscosity models (LEVMs) such as the k – ε and k – ω types are still most widely chosen for their robustness and computational efficiency [3]. However, LEVMs have well-known assumptions that limit their predictive accuracy when simulating certain types of flows. A major example is the Boussinesq Hypothesis (BH), which assumes that the Reynolds stress tensor is linearly proportional to the mean strain rate tensor. This simplification is invalid for a wide range of flows including those with separation or secondary motion [1]. While other RANS model classes such as nonlinear eddy viscosity models (NLEVMs) and Reynolds stress models have been developed with more advanced theoretical bases, they lack robustness and can only predict certain flows more accurately [4].

Pursuits in improving RANS approaches and recent rapid developments in machine learning (ML) have motivated tremendous growth in data-driven RANS turbulence modelling [5]. Due to the popularity of LEVMs, significant efforts have been directed towards developing data-driven models that predict the Reynolds stress tensor or one of its representations more accurately than BH [6]. These include its Eigendecomposition [6, 7], its divergence [8, 9], the optimal eddy viscosity [10, 11], and the normalised Reynolds stress anisotropy tensor, which will be referred to as the anisotropy tensor hereafter [12, 13]. The tensor basis neural network (TBNN) is an early example which inspired most other works in this field and falls into the latter category [14].

While such models are often shown to give more accurate predictions than traditional RANS approaches, the results are usually based on test cases that have similar flow features to the training cases [15]. In contrast, it is widely reported in the literature that the prediction accuracy of these models can deteriorate otherwise [16]. This reliability issue can cast doubt on model accuracy during deployment, as there are no target results at this final stage to assess model predictions [17]. While some reliability doubt may be acceptable in research environments, models are required to give results that are accurate across different flows in some industrial applications, hence assurance of model reliability would be important. As these models have begun being applied to flows resembling practical cases including nuclear [18, 19], turbomachinery [20], cooling [21, 22], and urban wind fields [23], more efforts from the community now ought to be focused on addressing this issue. Although doubt may be alleviated by using a universally accurate data-driven model, the possibility of its existence is still an open question [24].

To provide risk-informed predictions, a data-driven model that contains uncertainty quantification (UQ) capabilities may be used. Several methods have already been proposed for RANS closure [25]. Some are based on a Bayesian formalism, whereby a prior distribution is specified over the model parameter values, and the posterior distribution over the parameters are computed during training [26]. For example, Geneva and Zabaras [27] and Tang et al. [28] developed Bayesian neural network (BNN) variants of the TBNN, which directly treat its parameters as stochastic variables with probability distributions that are learnable using Bayesian inference. However, even with common approximate methods such as variational inference, training BNNs can be significantly more computationally expensive than standard non-Bayesian NNs [29]. This becomes especially problematic as the datasets and number of model parameters grow [30]. Some approaches have been proposed to impose a hierarchical prior over the hyperparameters of the model parameter prior distributions. For example, Agrawal and Koutsourelakis [31] applied Bayesian inference with hierarchical priors to a TBNN, and Scillitoe et al. [32] trained Mondrian forests to predict the Eigendecomposition of the Reynolds stress tensor. Similarly, Cherroud et al. [33] and Carlucci et al. [34] extended Schmelzer et al.'s sparse symbolic regression framework [13] to include Bayesian learning with hierarchical Laplace priors. As the choice of priors can greatly influence the model posterior results, it is common practice to conduct a prior sensitivity analysis [35]. However, this process can also be computationally expensive, as multiple models are required to be trained and each involves repeated re-estimation of hierarchical priors [26].

Non-Bayesian data-driven RANS models with UQ have also been proposed in the literature. For example, Man et al. [36] used ensemble TBNNs, where the parameters of each member network have different random initialisations and dataset shuffles. Ensemble predictions can then be represented with a mean and variance parameterising the overall uncertainty [29]. This type of analysis can also be undertaken with the ensemble Kalman method proposed by Zhang et al. [37]. Although ensemble methods are straightforward to implement and easily parallelisable, they require high computational memory if many member networks with a considerable number of parameters are used [38]. While moderate-sized networks have mostly been trained for data-driven RANS modelling, it is expected that with the growing availability of datasets from complex flow cases, very large networks will be required to model various nonlinearities across different turbulent flow regimes accurately [39]. Hence, the computational effort and memory requirements in data-driven turbulence modelling necessitate alternative UQ methods to be explored.

In the wider context of data-driven computational fluid simulations with UQ, some recent studies have explored the use of mean-variance estimation networks (MVENs), such as for reduced-order modelling [40], and LES sub-grid scale modelling [41]. MVENs are deterministic neural networks that directly predict the mean and standard deviation of chosen output variables. Therefore, they are built on the premise that the output can be parameterised by these statistical quantities, which facilitate uncertainty predictions [42]. MVENs can be simply trained with maximum likelihood estimation, and uncertainty can be easily computed with a simple forward pass in a single deterministic network [38]. This makes MVENs significantly more efficient to train than existing models based on Bayesian inference, and less memory consuming than ensemble methods [35].

In this paper, we address the challenge of reliability in existing data-driven RANS turbulence models by introducing a novel approach that uses mean-variance estimation networks (MVENs). The core objective of this work is to enhance RANS turbulence modelling with efficient uncertainty and error quantification by integrating an MVEN into a tensor basis neural network (TBNN). The proposed framework not only preserves the predictive accuracy of existing data-driven models but also incorporates insightful uncertainty and error quantification. Our contribution lies in demonstrating how MVENs can provide accurate predictions alongside meaningful uncertainty metrics, offering a more comprehensive understanding of turbulence behaviour. This work sets a foundation for future advancements in data-driven turbulence modelling, promoting greater reliability and applicability across diverse flow scenarios. The efficient uncertainty and error quantification functionalities represent a significant step forward in addressing the limitations of existing data-driven RANS models, enabling risk-informed predictions for a wide range of engineering applications.

## 2 Background

### 2.1 RANS Equations and the Reynolds Stress Tensor

Turbulent flows are commonly simulated using Reynolds-averaged Navier Stokes (RANS) approaches due to their computational efficiency [1]. These methods involve solving the RANS equations, which are given as the following in incompressible, isothermal, and steady state form:

$$\frac{\partial \bar{u}_i}{\partial x_i} = 0$$

$$\frac{\partial \bar{u}_i \bar{u}_j}{\partial x_j} = -\frac{1}{\rho}\frac{\partial \bar{p}}{\partial x_i} + \nu \frac{\partial}{\partial x_j}\left(\frac{\partial \bar{u}_i}{\partial x_j} + \frac{\partial \bar{u}_j}{\partial x_i}\right) - \frac{\partial \tau_{ij}}{\partial x_j} \quad (1)$$

where $\bar{u}_i, \bar{p}, x_i, \nu$, and $\tau_{ij}$ ($= \overline{u'_i u'_j}$) are the mean velocity, mean pressure, spatial coordinates, kinematic viscosity, and Reynolds stress tensor, respectively [3]. The Reynolds stress tensor represents the average rate of momentum transfer caused by turbulent fluctuations and must be modelled to close the RANS equations [43]. This is usually undertaken by decomposing the Reynolds stress into its isotropic and anisotropic parts:

$$\overline{u'_i u'_j} = 2k b_{ij} + \frac{2}{3}k \delta_{ij} \quad (2)$$

where $k$, $b_{ij}$, and $\delta_{ij}$ represent the turbulent kinetic energy (TKE), anisotropy tensor, and Kronecker delta, respectively. For brevity, $b_{ij}$ will also be written as $\boldsymbol{b}$ herein. The isotropic $2k\delta_{ij}/3$ part is absorbed into the pressure term of Eq. (1), leaving the anisotropic $2k\boldsymbol{b}$ part to be modelled [44]. In typical two-equation RANS approaches such as the k – ε and k – ω types, $\boldsymbol{b}$ is modelled under the assumption that it is directly proportional to the mean strain rate with the Boussinesq Hypothesis (BH). While this is reasonable for simple shear flows, it can give inaccurate predictions for a wide range of other flows, including those with separation or secondary motion [43].

### 2.2 The Generalisability Problem

In search of other methods to predict the anisotropy tensor $\boldsymbol{b}$ more accurately, numerous data-driven RANS models have been developed and proposed in the literature [30]. These typically demonstrate promising results, as more accurate predictions than traditional RANS models are often shown [5]. However, their comparisons are usually based on test cases that contain the same or similar flow features as the training cases. When data-driven RANS models are tested on cases that have different flow phenomena to the training cases, they are often reported to give inaccurate predictions that can be worse than traditional RANS model results [24]. This lack of generalisability has persistently limited the reliability of data-driven RANS models [15, 27, 45].

Studies often cite dataset shift as the causes of generalisability error [46, 47]. Dataset shift is the notion that a supervised model trained on a limited dataset may learn input-to-output patterns that are invalid when it is evaluated on the test dataset [48]. This can be addressed by using a richer training dataset that encompasses various turbulent regimes, which enables the model to learn a broader range of patterns [39]. Using a sufficient number of relevant physics-informed input features that are non-dimensional and bounded appropriately has also been discussed to promote generalisability [49]. These can enable overlap in the input space and allow one-to-one input-to-output mappings to be made. Otherwise, error due to non-unique mapping can occur, where similar input values require different model responses for different flow phenomena [24, 50].

Both causes can be addressed by using a universal generalised model that is able to predict a diverse range of turbulent regimes accurately. However, it is expected that a very complex form (*e.g.*, many hidden layers and nodes in a NN) will be required to model different input-to-output relationships for various bifurcations and nonlinearities [39]. Using simpler expert models that constitute a universal data-driven model would be a more computationally affordable alternative, where each expert model is optimised on a specific turbulent regime. Such approaches have been proposed in the literature with user-specified marker functions [51], decision tree classification [52], and mixture of experts [53, 54] to detect different flow phenomena. However, questions arise regarding how many expert models should be trained, and how their predictions should be combined [24].

The above discussion shows that the development of universal data-driven RANS models is still in its infancy, while there are doubts whether one can exist [39]. Therefore, the reliability of existing data-driven RANS models remains in question into the foreseeable future. The proposed data-driven model in the present work aims to embed reliability metrics into its predictions, thereby enabling risk-informed results to be given. A holistic view of the generalisability problem will be taken, such that generalisability error shall be treated and quantified as a combined consequence of dataset shift and inadequate input features.

## 3 Methodology

### 3.1 Mean-Variance Estimation Networks

The UQ and EQ capabilities of the proposed model is enabled by an MVEN architecture and methodology. MVENs are deterministic NNs that directly predict the mean $\boldsymbol{\mu}$ and standard deviation $\boldsymbol{\sigma}$ of the targets $\boldsymbol{y}$ at the output nodes as illustrated in **Figure 1** [55]. Any MVEN denoted $\mathcal{M}$ can be generally represented as $\mathcal{M}: \boldsymbol{x} \mapsto (\boldsymbol{\mu}, \boldsymbol{\sigma})$, where $\boldsymbol{\mu}$ and $\boldsymbol{\sigma}$ are taken to be continuous functions of input $\boldsymbol{x}$. While $\boldsymbol{x}$ is usually a vector, $\boldsymbol{y}$ and its statistical parameters $\boldsymbol{\mu}$ and $\boldsymbol{\sigma}$ may be scalars or vectors. Although $\boldsymbol{\mu}$ and $\boldsymbol{\sigma}$ were predicted by separate sub-networks in the original MVEN [42], an alternative approach is taken herein where $\boldsymbol{\mu}$ and $\boldsymbol{\sigma}$ are predicted by the same hidden layers. This approach is based on a mixture density network (MDN), which is a class of NNs that model the output as a mixture of probability density functions (PDFs) with the assumption that predictions are generated from a distribution [56]. By modelling the output of an MDN as a single PDF instead of a mixture of PDFs, the MDN becomes an MVEN as a special case. This MVEN implementation has also been applied to other computational fluid problems [40, 41].

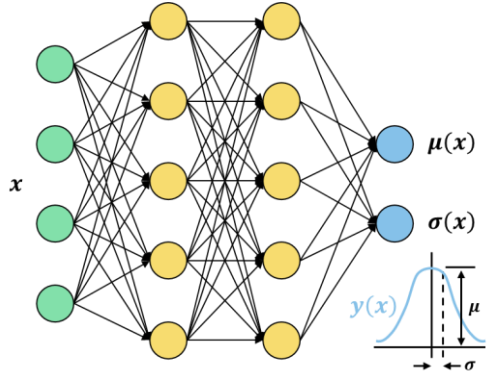

**Figure 1** An example of a mean-variance estimation network with the single probability density function mixture density network approach.

It can be shown with statistical learning theory that the standard deviation $\boldsymbol{\sigma}$ predicted by an MVEN is linearly proportional to the root mean squared error (RMSE) from a corresponding standard NN, provided that both models have been fully trained with infinite data that spans the entire input $\boldsymbol{x}$ and target $\boldsymbol{y}$ space [57]. In practice, it is impossible to reach this infinite training data limit due to computational limitations. However, this proportionality relationship can be extended to suggest that converged $\boldsymbol{\sigma}$ predictions from an MVEN could be a proxy variable for the converged RMSE from a corresponding standard NN [26]. The derivation of this relationship can be found in **Appendix A**. If the mean $\boldsymbol{\mu}$ predictions given by the MVEN are similar to the predictions of the standard NN, then a single MVEN can predict means $\boldsymbol{\mu}$ while its converged RMSE can be estimated with its $\boldsymbol{\sigma}$ predictions. In fact, results of the current work show that there is approximate proportionality. Therefore, $\boldsymbol{\sigma}$ predictions given by an MVEN may be especially insightful during model deployment, as they can be used for the EQ of predictions in the absence of target results to make risk-informed decisions. In the context of turbulence modelling, generalisability error may thereby be easily identified.

### 3.2 TenMaven: Tensor Basis Mean and Variance Estimation Neural Network

To demonstrate the capabilities of MVEN in data-driven RANS turbulence modelling, a new model called the tensor basis mean and variance estimation neural network (TenMaven) is introduced in this paper. The TenMaven shown in **Figure 2** builds upon Ling et al.'s [14] original tensor basis neural network (TBNN) by integrating an MVEN in the architecture for UQ and EQ purposes. As with the original TBNN, the TenMaven is based on the general effective viscosity hypothesis (GEVH), and therefore also predicts anisotropy tensor $\boldsymbol{b}$ as a Galilean invariant function of the non-dimensional mean strain rate $S$ and mean rotation rate $R$ under local equilibrium [58]. These are defined as $S = t_S S$, and $R = t_R R$, where $t_S$ and $t_R$ represent timescales for the dimensional mean strain rate $S \left(= (1/2)[(\partial \bar{u}_i/\partial x_j) + (\partial \bar{u}_j/\partial x_i)]\right)$ and dimensional mean rotation rate $R \left(= (1/2)[(\partial \bar{u}_i/\partial x_j) - (\partial \bar{u}_j/\partial x_i)]\right)$, respectively. For normalisation purposes, $t_S$ and $t_R$ are set to the commonly used normalising timescales $1/(\|S\| + (\varepsilon/k))$ and $1/(2\|R\|)$ respectively, which were introduced by Wu et al. [10] where brackets $\|\cdot\|$ represent the Frobenius norm of the quantity in them, and $\varepsilon$ denotes the TKE dissipation rate. Under this assumption that $\boldsymbol{b} = \boldsymbol{b}(\boldsymbol{S}, \boldsymbol{R})$, The GEVH states that $\boldsymbol{b}$ can be fully determined by a finite sum of ten basis tensors in the most general incompressible case due to the Cayley-Hamilton theorem:

$$\boldsymbol{b} = \sum_{n=1}^{10} g_n(\boldsymbol{\lambda}) \, \boldsymbol{T}_n \tag{3}$$

where $g_n$ are coefficient scalars and $\boldsymbol{T}_n$ are tensor products of $\boldsymbol{S}$ and $\boldsymbol{R}$ given in **Appendix B** [58]. These will be represented collectively as $\boldsymbol{g} = [g_1, \ldots, g_{10}]^T$ and $\boldsymbol{\mathcal{T}} = [\boldsymbol{T}_1, \ldots, \boldsymbol{T}_{10}]$, respectively hereafter. The task becomes predicting $\boldsymbol{g}$ in a pointwise manner, which can be performed with a data-driven model such as the TBNN or TenMaven.

Pope [58] supposed the $\boldsymbol{g}$ coefficients are dependent on the following scalar invariants of the minimal integrity basis formed by $\boldsymbol{S}$ and $\boldsymbol{R}$: $\lambda_1 = tr(\boldsymbol{S}^2)$, $\lambda_2 = tr(\boldsymbol{R}^2)$, $\lambda_3 = tr(\boldsymbol{S}^3)$, $\lambda_4 = tr(\boldsymbol{R}^2\boldsymbol{S})$, and $\lambda_5 = tr(\boldsymbol{R}^2\boldsymbol{S}^2)$. The input dependencies will be represented generally as $\boldsymbol{\lambda}$, such that $\boldsymbol{\lambda} = \{\lambda_1, \ldots, \lambda_5\}$ in this postulation. While many tensor basis models proposed in the literature only use these invariants, including those formed by pressure gradient $\nabla p$ and TKE gradient $\nabla k$ in the $\boldsymbol{\lambda}$ set has become increasingly common among recent models to account for non-equilibrium turbulence effects [20, 59]. Furthermore, Man et al. [50] showed that non-unique mapping can easily occur when only the scalar invariants formed by $\boldsymbol{S}$ and $\boldsymbol{R}$ are selected in the $\boldsymbol{\lambda}$ set. Some models have been developed with supplementary heuristic scalars in the $\boldsymbol{\lambda}$ set to consider additional physical information as well [16, 60]. These trends are adopted in the current work, where 55 scalar input features shown in **Appendix C** were used, such that $\boldsymbol{\lambda} = \{\lambda_1, \ldots, \lambda_{55}\}$. The first 47 scalar inputs are invariants formed by $\boldsymbol{S}$, $\boldsymbol{R}$, $\nabla p$, and $\nabla k$ introduced by Pope [58] and Wu et al. [10] resulting from the Cayley-Hamilton Theorem. The remaining 8 are heuristic scalars that were chosen based on their high feature importance scores in various studies. More details on the scalar inputs can be found in **Appendix C**, including their normalisation.

In any case of chosen input features $\boldsymbol{\lambda}$, data-driven RANS models based on GEVH are developed with the aim of approximating the functional mapping $\mathcal{F}: \boldsymbol{\lambda} \mapsto \boldsymbol{g}$ that can give accurate predictions of the anisotropy tensor $\boldsymbol{b}$. This is achieved by fitting the model parameters (*e.g.*, the weights and biases in NNs) which occurs during the training process. In the TBNN and TenMaven, these are found between the scalar input layer $\boldsymbol{\lambda}$ and the final hidden layer $\boldsymbol{z}$ as shown in **Figure 2**. Predicted $\boldsymbol{g}$ coefficients in the final hidden layer $\boldsymbol{z}$ are then multiplied by their corresponding tensor in $\boldsymbol{\mathcal{T}}$, and the resulting scaled tensors are summed to give anisotropy tensor $\boldsymbol{b}$ predictions to match Eq. (3).

Compared to the original TBNN, the proposed TenMaven model includes the following enhancements for UQ and EQ estimation. First, instead of predicting the $\boldsymbol{g}$ coefficients and subsequent anisotropy tensor $\boldsymbol{b}$ discretely, their means are predicted instead as shown in **Figure 2** with the use of $\boldsymbol{\mu_g}$ and $\boldsymbol{\mu_b}$. Second, mean values of the anisotropy tensor $\boldsymbol{\mu_b}$ components are considered to have uncertainty intervals centred on them. It is assumed that given inputs $\boldsymbol{\lambda}$, the average uncertainty over all components of $\boldsymbol{\mu_b}$ can be represented by a scalar standard deviation $\sigma_b$ denoted in non-bold. Although this simplification is adopted in the current work to follow the MDN approach of Bishop [56], covariance matrices may be predicted for each component at the cost of a more complex formalism. The standard deviation $\sigma_b$ is predicted by the $z_\sigma$ node in the final hidden layer with an exponential output activation function to ensure positive predictions. A scaling factor of 10 was found to reduce model overfitting. Third, to be consistent with the MDN approach, another node denoted by $z_\alpha$ has been included in the final hidden layer. Its purpose is to predict the mixing probability of each PDF in an MDN prediction. Given that only one PDF is used in the TenMaven, $z_\alpha$ and $\alpha$ are always predicted as 1. As with the mean coefficient nodes $\boldsymbol{\mu_g}$, both $z_\sigma$ and $z_\alpha$ are also fully connected to the hidden layers.

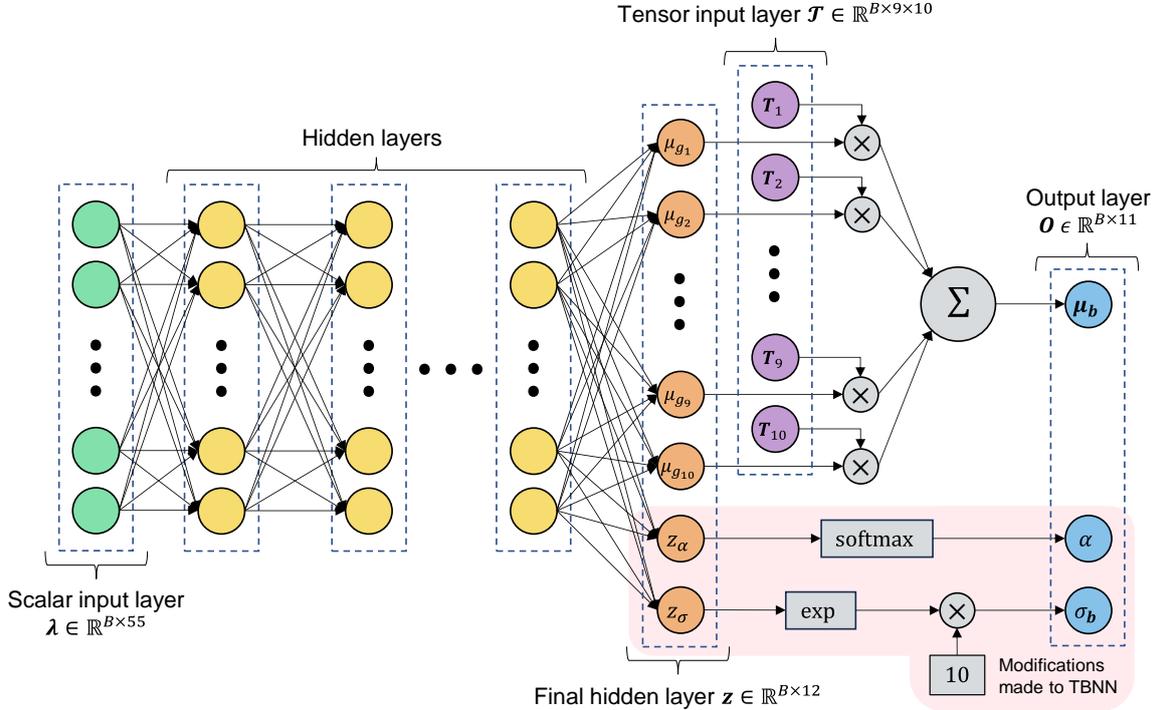

**Figure 2** Schematic of the proposed TenMaven model architecture. Differences between the TenMaven and TBNN by Ling et al. [14] are highlighted in light red.

In related work, Grogan et al. [61] trained an ensemble of MVENs to predict the mean and variance of $\boldsymbol{g}$ coefficients in an algebraic Reynolds stress model applied to two-dimensional homogeneous turbulence. The current work aims to enhance typical data-driven RANS models such as the TBNN with mean-variance estimation, which are commonly trained to predict significantly more complex flow physics, including flow separation and secondary motions. The TBNN was chosen in this demonstration over other data-driven models as it has been impactful and commonly used. Therefore, the MVEN enhancements introduced herein can be easily implemented in many existing codes. Furthermore, many other data-driven models are based on the GEVH, including tensor basis model alternatives [62, 63], and symbolic regression models [12, 13]. Although some of these approaches are not based on a NN, it is expected that similar methodologies to the TenMaven for mean-variance estimation can be devised.

### 3.3 Training Process and Loss Function

The TBNN and TenMaven are trained by a supervised machine learning (ML) algorithm on a training dataset. During the training process, entries of the training dataset are randomly shuffled and typically divided into $N_{batch}$ minibatches [17]. Afterwards, each minibatch is fed into the model and forward-propagation is executed to give predictions at the output nodes. The predictions are then compared with prescribed target values to compute a residual error called a loss. This loss is backpropagated through the model, which allows an optimiser to update the model parameters (*i.e.*, the weights and biases). One complete execution of forward- and back-propagations for all minibatches is known as an epoch. The number of epochs increases until a convergence criterion to terminate the training process is satisfied. For models based on the GEVH such as the TBNN and TenMaven, the training and all other datasets are comprised of i) scalars $\boldsymbol{\lambda}$ and tensors $\boldsymbol{\mathcal{T}}$ calculated from pointwise RANS simulation results of distinct flow cases, and ii) target anisotropy tensor $\boldsymbol{b}$ values given by corresponding scale-resolved simulations of the cases at the same pointwise locations [46]. The scalars $\boldsymbol{\lambda}$ and tensors $\boldsymbol{\mathcal{T}}$ of each minibatch are passed into their respective input layers shown in **Figure 2** to begin forward-propagation, and the loss is computed by comparing the $\boldsymbol{b}$ predictions with the target $\boldsymbol{b}$ values.

Compared to multilayer perceptrons which have been commonly used for data-driven RANS modelling, two additional considerations must be taken when implementing an MVEN. First, a distribution that describes how the predicted uncertainty is centered on the mean must be chosen for loss calculation. The Gaussian distribution is used in the present work such that $\boldsymbol{b} \sim \mathcal{N}(\boldsymbol{\mu_b}(\boldsymbol{\lambda}), \sigma_b(\boldsymbol{\lambda}))$, as this allows loss functions based on maximum likelihood estimation (MLE) to be easily calculated [64]. Although mean squared error (MSE) loss is typically used in regression problems, its inability to assess uncertainty necessitates the use of loss functions that are based on MLE when training an MVEN. This leads to the second consideration of choosing an MLE-based loss function. Mean negative log likelihood (MNLL) loss $E$ was used as i) taking the logarithm of the likelihood avoids potential precision underflow and ii) taking the negative turns loss maximisation into a minimisation, which is more intuitive for ML [65]. The MNLL loss in the current work can be written as

$$E = \frac{1}{2}\ln(2\pi) + \frac{1}{B}\sum_{q=1}^{B}\ln\sigma_{\boldsymbol{b}}^{(q)} + \frac{1}{18B}\sum_{q=1}^{B}\frac{1}{\sigma_{\boldsymbol{b}}^{(q)}\sigma_{\boldsymbol{b}}^{(q)}}\sum_{i=1}^{3}\sum_{j=1}^{3}\left(\mu_{b_{ij}}^{(q)} - b_{ij}^{(q,targ)}\right)^2 \quad (4)$$

where index $q = 1, \ldots, B$ denotes minibatch entries, and superscript $targ$ represents target values of $\boldsymbol{b}$. Full derivation of Eq. (4) can be found in **Appendix D**. The training process of the TenMaven is detailed in **Algorithm 1**, where index $t$ denotes the minibatch count. **Algorithm 1** can also be used as a basis to develop data-driven RANS models along with implementation of mean-variance estimation.

---

**Algorithm 1:** Training process for the TenMaven

1    Initialise model parameters $\boldsymbol{w}$
2    **while** mean negative log likelihood (MNLL) loss $E$ is not converged **do**
3      **for** each minibatch $t = 1, \ldots, N_{batch}$ in training dataset **do**
4        **for** each entry $q = 1, \ldots, B$ in $t$ **do**
5          Compute model predictions: $\left(\alpha^{(q)}, \boldsymbol{\mu}_{\boldsymbol{b}}^{(q)}, \sigma_{\boldsymbol{b}}^{(q)}\right) \leftarrow \mathcal{M}\left(\boldsymbol{\lambda}^{(q)}, \boldsymbol{\mathcal{T}}^{(q)}, \boldsymbol{w}\right)$
6        **end for**
7        Compute MNLL loss with Eq. (4): $E \leftarrow \left(\boldsymbol{\mu}_{\boldsymbol{b}}^{(t)}, \boldsymbol{\sigma}_{\boldsymbol{b}}^{(t)}, \boldsymbol{b}^{(t,targ)}\right)$
8        Update network parameters: $\boldsymbol{w} \leftarrow Optimiser(\nabla_{\boldsymbol{w}}E, \boldsymbol{w})$
9      **end for**
10   **end while**

---

A validation process may be executed after a chosen number of epochs. This involves evaluating the model on a validation dataset, which contains examples that the model did not encounter during training. By comparing the model's predictions against the target values of the validation dataset, a loss can be computed which gives an indication of the model's generalisation capability on unseen cases. The training process terminates when a convergence criterion such as early stopping is met [66]. This entails terminating when the average of the $N_{vl}$ most recent validation losses is larger than the average of the previous $N_{vl}$ validation losses where $N_{vl}$ is a pre-specified number, which prevents overfitting. If this criterion is not met after many epochs, a maximum number of epochs can be set to terminate training. In the present work, validation was undertaken after every 10 epochs, $N_{vl}$ was set to 5, and the maximum number of epochs was chosen as 10000 to ensure sufficient model convergence. Other hyperparameter settings that were chosen are shown in **Table 1**. After training is complete, the model can be evaluated on a held-out testing dataset to assess its expected accuracy on unseen cases during deployment. All ML processes were performed using PyTorch 1.13.

**Table 1** Hyperparameter settings

| Hyperparameter | Setting |
| --- | --- |
| Number of hidden layers | 5 |
| Number of nodes per hidden layer | 20 |
| Hidden node activation functions | Rectified Linear Unit |
| Optimiser | Adam |
| Learning rate | 0.0001 |
| Batch size, $B$ | 16 |

### 3.4 Datasets

The training dataset was composed of data from two flow over periodic hills cases with steepness factor $\beta = 0.5$ and 1.5 as shown in **Figure 3(a)**. These two-dimensional cases are characterised by the interaction of flow over repetitive hill-like structures, which includes near-wall boundary layer dynamics, separated flow, and recirculation. The flow enters the domain from the left with a bulk inlet velocity $U_b$ and separates over the left hill of height $H_h$. This creates a recirculation zone – the extent of which is dependent on the steepness factor $\beta$, and the flow subsequently reattaches downstream [67]. Similarly, the validation dataset and the first case in the test dataset were also based on flow over periodic hills with $\beta = 1.0$ and 1.2, respectively. All four cases have the same Reynolds number Re of 5600, which is dependent on $U_b$ and $H_h$. To evaluate the generalisation capabilities of the TenMaven on an entirely different flow, a second case in the test dataset was prepared using data from a cross-section of flow in a square duct case at Re = 3500, based on the bulk inlet velocity $U_b$ and duct half-height $H_d/2$. This case contains a well-known pattern of secondary flow in the cross-plane of the duct – namely pairs of counter-rotating vortices in each corner as shown in **Figure 3(b)**.

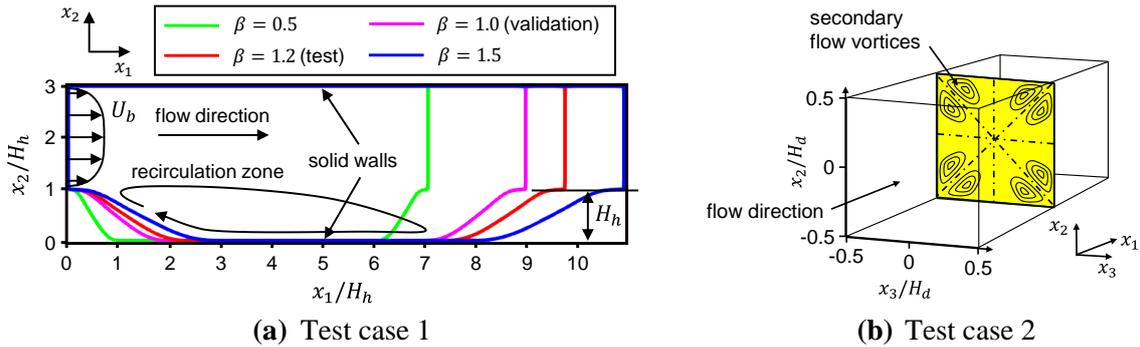

**(a)** Test case 1   **(b)** Test case 2

**Figure 3** Schematic of the present test cases including **(a)** flow over periodic hills cases with steepness factors $\beta = 0.5, 1.0, 1.2,$ and 1.5 at Re = 5600 based on bulk inlet velocity of 0.028 m/s and hill crest height $H_h$ of 1m, and **(b)** square duct case at Re = 3500 based on bulk inlet velocity of 0.482 m/s and duct half-width of 0.5m.

These cases were chosen as their flow phenomena are known to pose challenges for RANS turbulence models based on BH [7]. For flow separation as featured in the periodic hill cases, such models typically underpredict the magnitudes of TKE and shear anisotropy component $b_{12}$. These lead to significant underprediction in the magnitude of Reynolds shear stress $\tau_{12}$ through Eq. (2), whereas predicting $\tau_{12}$ accurately is essential for obtaining the correct mean recirculation field and reattachment length [51]. For secondary flows such as the corner vortices of the square duct case, accurate predictions are dependent on the correct anisotropic imbalance between normal Reynolds stress components. However, BH cannot predict secondary flows at all, as it enforces an isotropic assumption with the third term in Eq. (2) [3]. Therefore, evaluating the TenMaven on these cases may enable its merits over BH to be highlighted clearly.

For each case, the tensors $\mathcal{T}$ shown in **Appendix B** and scalars $\lambda$ listed in **Appendix C** were computed using data from RANS shear stress transport (SST) simulations of the cases performed by McConkey et al. [68]. Target anisotropy tensor **b** values were obtained from time-averaged DNS of the periodic hill cases and the square duct case undertaken by Xiao et al. [67] and Pinelli et al. [69], respectively. All quantities correspond to the cell centres of the RANS mesh, resulting in 14751 data points (*i.e.*, dataset entries) for each periodic hill case, and 9216 for the square duct case.

## 4 Results

To compare the TenMaven results against a well-established data-driven model, a TBNN was also trained and tested on the same cases with the same inputs and hyperparameter settings. As the TBNN does not predict uncertainty, MSE between predicted anisotropy $\hat{b}_{ij}^{(q)}$ and target anisotropy $b_{ij}^{(q,targ)}$ over all **b** components and minibatch entries was used as the loss function instead, given as $\left(\sum_{q=1}^{B}\sum_{i=1}^{3}\sum_{j=1}^{3}\left(\hat{b}_{ij}^{(q)}-b_{ij}^{(q,targ)}\right)^{2}\right)/9B$.

**Table 2** shows training and validation metrics gathered from the TBNN and TenMaven. The TBNN converged in 870 epochs with a training time of 7.8s per epoch, resulting in a total run time of 2 hours. In comparison, the TenMaven only required 220 epochs to converge with a training time of 14.5s per epoch, resulting in a shorter total run time of 53 minutes. Therefore, the TenMaven is shown to require approximately double the amount of computation per epoch but may converge sooner than the TBNN. These training times represent an upper limit as only a single CPU was used, whereas using more processors is expected to reduce the computation time. The final average training and validation losses of both models are also shown in **Table 2**. Note that negative values are possible for negative log likelihood losses, where reducing them is still the aim.

**Table 2** Training and validation metrics

|  | TBNN | TenMaven |
|---|---|---|
| Number of epochs | 870 | 220 |
| Training time per epoch | 7.8 seconds | 14.5 seconds |
| Total training time | 2 hours | 53 minutes |
| Final average training loss per batch | $2.65\times10^{-4}$ (MSE) | -3.24 (MNLL) |
| Final average validation loss per batch | $4.47\times10^{-4}$ (MSE) | -2.66 (MNLL) |

To assess the predictive accuracy of both models, the $\boldsymbol{b}$ and $\boldsymbol{\mu_b}$ predictions given by TBNN and TenMaven respectively, on the test cases are compared against RANS SST and DNS results in the following analysis. For brevity, the $\boldsymbol{\mu_b}$ results will simply be referred to as $\boldsymbol{b}$.

## 4.1 Test Case 1: Flow over Periodic Hills

### 4.1.1 Anisotropy Tensor Predictions

Predictions of the anisotropy tensor $\boldsymbol{b}$ components given by RANS SST, DNS, TBNN, and TenMaven for the periodic hills test case are shown in **Figure 4**. A comparison between the RANS and DNS results shows that RANS cannot predict the normal components accurately in all regions of the domain. The poor predictions of $b_{11}$ and $b_{22}$ result from the use of Boussinesq hypothesis (BH), which calculates Reynolds stress by modelling the anisotropy tensor as $\boldsymbol{b} = -C_\mu k S/\varepsilon$. This expression only gives valid predictions of $\boldsymbol{b}$ in flows where alignment between the mean strain rate tensor $S$ and $\boldsymbol{b}$ exists, whereas there is misalignment between the normal components of these tensors in flow over periodic hills cases [10]. RANS by virtue of BH gives zero for $b_{33}$ entirely, as velocity gradient $\partial \bar{u}_3 / \partial x_3$ is predicted zero across the whole domain due to two-dimensionality.

In contrast, both ML models predict the normal components of $\boldsymbol{b}$ significantly more accurately than RANS throughout the entire domain due to their increased flexibility in modelling $\boldsymbol{b}$. Rather than constraining $\boldsymbol{b}$ to a linear alignment with $S$, the GEVH given by Eq. (3), the choice of scalars in $\boldsymbol{\lambda}$, and nine extra tensors in $\boldsymbol{\mathcal{T}}$ allow the ML models to account for higher strain, rotation, anisotropic, and non-equilibrium effects in modelling $\boldsymbol{b}$. These improvements are most noticeable in the near-wall, separation, and recirculation regions. Furthermore, although $\boldsymbol{\mathcal{T}}$ was calculated from two-dimensional RANS results, non-zero values of $b_{33}$ could be predicted by the ML models as $b_{33}$ is dependent on non-zero components in $\boldsymbol{\mathcal{T}}$. These predictions are even in very good agreement with DNS, which demonstrates that data-driven RANS models based on the GEVH are able to predict turbulence in the third dimension for two-dimensional flows. A comparison between the normal component predictions given by TBNN and TenMaven shows that more noise exists in the TBNN results. This may be caused by overfitting as a result of the MSE loss optimising the TBNN to improve $\boldsymbol{b}$ predictions exclusively, whereas optimising additionally for $\sigma_b$ in the MNLL loss is believed to have provided a regularising effect in the TenMaven.

To quantitatively compare the $\boldsymbol{b}$ predictions against DNS results, the root mean squared error (RMSE) of each component was calculated as

$$RMSE_{ij} = \sqrt{\frac{1}{N} \sum_{q=1}^{N} \left( \hat{b}_{ij}^{(q)} - b_{ij}^{(q,targ)} \right)^2} \qquad (5)$$

where $N$ is the total number of datapoints (*i.e.*, dataset entries) for the test case, and $q = 1, ..., N$ denotes each datapoint. The hat notation $\widehat{\phantom{x}}$ represents model predictions given by RANS, TBNN, or the TenMaven. **Figure 5** presents the $RMSE_{ij}$ calculated for each component and method in the periodic hills test case, with accuracy improvement given by the ML models compared to RANS displayed in brackets. The TBNN and TenMaven are shown to have predicted the normal components compared to RANS significantly more accurately by 72-86% and 55-73% respectively on average, which supports the qualitative observations given above. It is believed that TBNN gave slightly higher accuracy than the TenMaven due to overfitting in the TBNN as previously discussed.

Although the normal components of Reynolds stress are important in many flows such as the second test case, it is the shear component $\tau_{12}$ that governs the behaviour of separated and recirculating flows [67]. **Figure 4** shows that the shear anisotropy component $b_{12}$ predicted by RANS is in satisfactory agreement with DNS throughout the entire domain due to approximate alignment between $S_{12}$ and $b_{12}$. However, some underprediction in magnitude can be found in the separation and recirculation regions, which is supported by previous findings in the literature [62]. Predicting $b_{12}$ accurately is necessary to give accurate calculations of $\tau_{12}$ and subsequent prediction of the reattachment length [51]. **Figure 4** qualitatively shows that TBNN and TenMaven are able to predict $b_{12}$ more accurately than RANS, especially in the flow contraction, separation, and recirculation regions. This is attributed to their greater flexibility in modelling ***b***, which resulted in $RMSE_{12}$ improvements of 65% and 55%, respectively on average for the two ML models across the whole domain as shown in **Figure 5**. Noise due to suspected overfitting is also shown in the TBNN prediction of $b_{12}$, which led to slightly higher accuracy compared to the TenMaven.

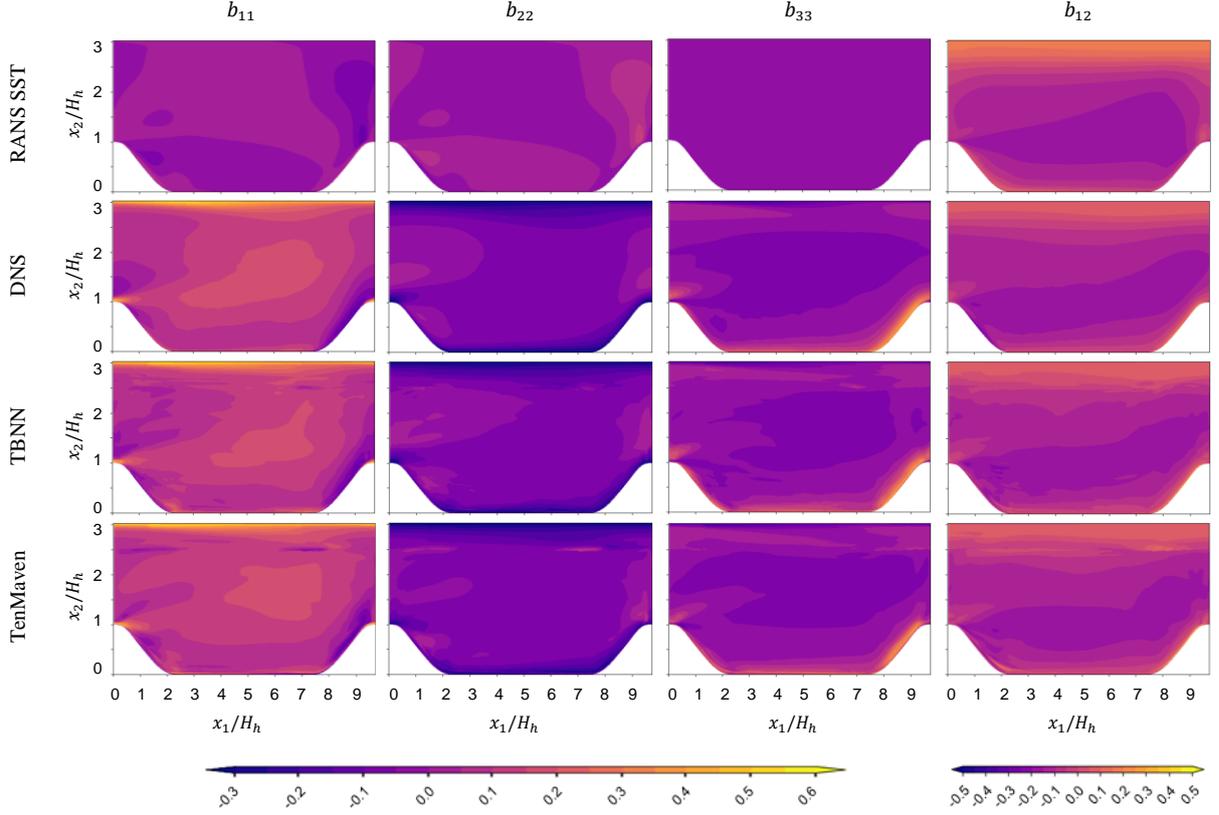

**Figure 4** Anisotropy tensor **b** component predictions of the periodic hills test case given by RANS SST, DNS, TBNN, and the TenMaven. Components $b_{13}$ and $b_{23}$ are not shown as their entire fields are zero. Minimum and maximum values of the colour bars are set close to the limits of realizability: $-1/3 \leq b_{ii} \leq 2/3$ and $-1/2 \leq b_{ij} \leq 1/2$, such that $i \neq j$ and $i,j \in \{1,2,3\}$.

To assess the overall accuracy of the models, the RMSE across all datapoints and **b** components was calculated as

$$RMSE_{all} = \sqrt{\frac{1}{9N}\sum_{q=1}^{N}\sum_{i=1}^{3}\sum_{j=1}^{3}\left(\hat{b}_{ij}^{(q)} - b_{ij}^{(q,targ)}\right)^2} \quad (6)$$

and are presented in the last column of **Figure 5** for each method. Compared to RANS, both TBNN and TenMaven are shown to have predicted **b** more accurately overall by 79% and 65% respectively, which are significant improvements. As mentioned earlier, the slight increase in accuracy given by TBNN compared to the TenMaven is believed to be caused by overfitting in the TBNN.

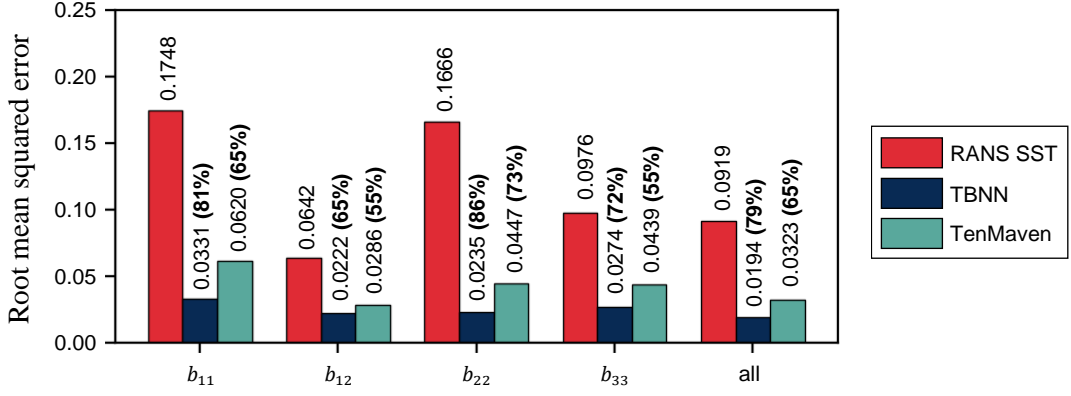

**Figure 5** Root mean squared error of anisotropy tensor $\boldsymbol{b}$ component predictions compared to DNS for the periodic hills test case. Accuracy improvement (%) = $(RMSE_{RANS} - RMSE_{ML}/RMSE_{RANS}) \times 100\%$ given by the ML models are shown in brackets.

### 4.1.2 TBNN and TenMaven Prediction Similarity

While **Figure 4** shows that the $\boldsymbol{b}$ predictions given by TBNN and the TenMaven are qualitatively similar, a quantitative evaluation was undertaken, as it was expected that this is a necessary condition for predicted standard deviation $\sigma_{\boldsymbol{b}}$ to act as a proxy variable for the error in the $\boldsymbol{b}$ predictions. **Figure 6** shows scatterplots of $\boldsymbol{b}$ components predicted by TBNN and TenMaven plotted against each other. Although some points deviate from the identity line, a strong positive linear correlation along it can be observed for all components. To quantify correlation for each component, the concordance correlation coefficient (CCC) $= 2\rho_p \sigma_x \sigma_y / \left(\sigma_x^2 + \sigma_y^2 + (\mu_x - \mu_y)^2\right)$ was calculated, where $\mu_x$ and $\mu_y$ are the means of the two sets of model results, $\sigma_x^2$ and $\sigma_y^2$ are the corresponding variances, and $\rho_p$ is the Pearson correlation coefficient [70]. The CCC measures how close the scatter points are to the identity line by assessing linearity and bias, where a value of 1 denotes perfect agreement between the two sets of results and a value less than 0.2 represents poor agreement [71]. **Figure 6** shows that high CCC values of 0.87 to 0.97 were calculated for the $\boldsymbol{b}$ components, thereby confirming strong similarity.

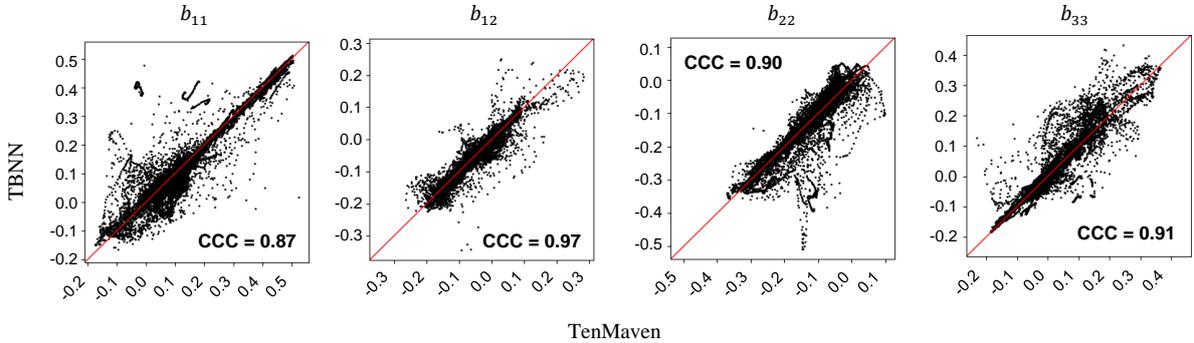

**Figure 6** Scatterplots of anisotropy tensor $\boldsymbol{b}$ components predicted by TBNN (y-axis) vs. TenMaven (x-axis) for the periodic hill test case. The red diagonal represents the identity line, where points that lie on it have been predicted identically by both models. The concordance correlation coefficient (CCC) between the two sets of predictions for each component are also shown.

### 4.1.3 Error Quantification

With strong similarity between the **b** predictions of TBNN and the TenMaven confirmed, the EQ given by predicted standard deviation $\sigma_b$ was assessed by examining its proportionality with the RMSE of the TenMaven **b** predictions. As the average $\sigma_b$ over all **b** components for each entry $q$ was predicted, the pointwise RMSEs averaged over all **b** components were calculated as

$$RMSE^{(q)} = \sqrt{\frac{1}{9}\sum_{i=1}^{3}\sum_{j=1}^{3}\left(\hat{b}_{ij}^{(q)} - b_{ij}^{(q,targ)}\right)^2} \qquad (7)$$

for comparison purposes, where $\hat{b}_{ij}^{(q)}$ is the pointwise mean anisotropy tensor predicted by TenMaven. **Figure 7(a)** shows a scatter plot of $\sigma_b$ plotted against $RMSE^{(q)}$. A strong positive correlation is demonstrated between them, which is supported by previous discussions about their proportionality. To assess its linearity, the Pearson correlation coefficient (PCC) was calculated which gives a value between -1 and 1, where the lower and upper bound indicate a perfect negative and positive linear correlation, respectively [72]. A high positive value of 0.88 was calculated for the PCC, which quantitatively supports the correlation observations made. An equation of $RMSE^{(q)} = 0.8\sigma_b$ was found for the least squares best fit line shown in red.

To relate the scatter points with their corresponding locations and to identify regions of high error, surface plots of $\sigma_b$ and $RMSE^{(q)}$ on the periodic hills domain are presented in **Figure 7(b)**. While some spikes in the $\sigma_b$ predictions were given, regions that have higher $RMSE^{(q)}$ values are shown to coincide with greater $\sigma_b$ predictions as expected. These are demonstrated to be present where low velocity gradients occur i) near the bottom wall and ii) in the inflection region along the $x_2/H_h = 2.6$ line as detailed by Mandler and Weigand [73] and Man et al. [50]. Due to the low velocity gradients, it was found that some component values of tensors $\mathcal{T}$ were significantly lower in these regions – approximately three orders of magnitude – than the rest of the domain. Therefore, it is believed that other tensors are required to be included in the input tensor set $\mathcal{T}$ to predict **b** accurately there, such as tensors involving pressure gradients proposed by Parmar et al. [74]. Nonetheless, **Figure 7** shows that $\sigma_b$ predicted by the TenMaven is an appropriate proxy variable for its error.

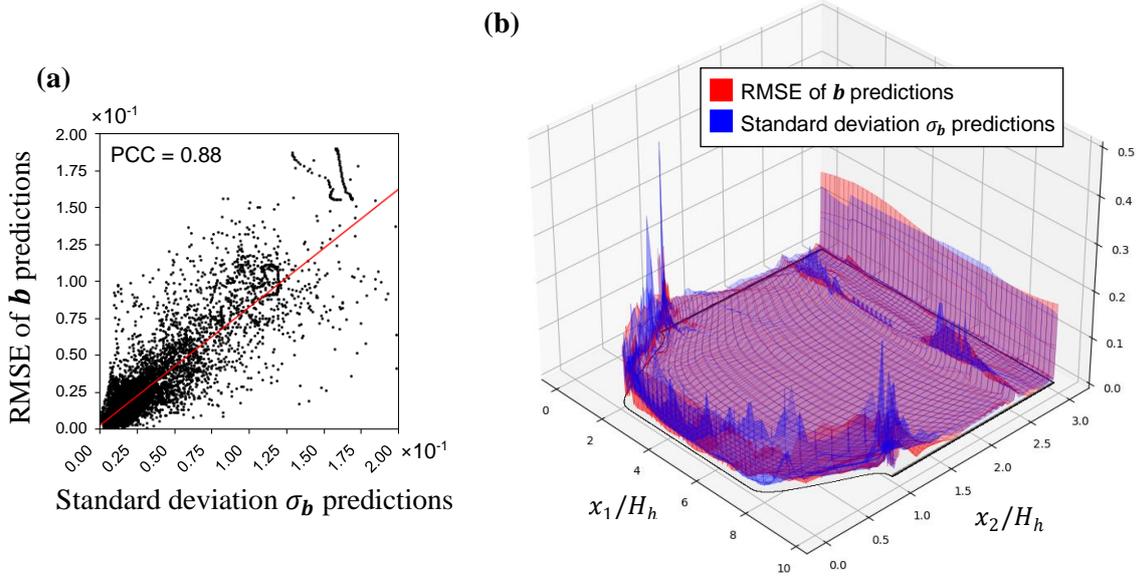

**Figure 7** RMSE of ***b*** predictions and standard deviation $\sigma_b$ predictions of the periodic hills test case given by TenMaven plotted **(a)** against each other on a scatter plot, and **(b)** as surfaces on the domain.

### 4.1.4 Uncertainty Quantification

To assess the UQ capability of $\sigma_b$ predictions, **Figure 8** shows line plots of $b_{12}$ with uncertainty bounds given by $\pm\sigma_b$ predicted by the TenMaven at three important streamwise locations: **(a)** immediately prior to separation, **(b)** at the centre of recirculation, and **(c)** at the reattachment point. These profiles are compared with $b_{12}$ predictions given by RANS, DNS, and TBNN. All plots show that the uncertainty bounds almost fully envelop the DNS profiles – even where the TenMaven deviates most from the DNS result. In regions where the TenMaven profiles are in very close agreement with DNS such as in the main flow of **Figure 8(c)** where $0.25 \leq x_2/H_h \leq 2.25$ and near the top wall where $2.5 \leq x_2/H_h \leq 3.0$, the $\sigma_b$ predictions are accordingly much lower.

The uncertainty bounds are found to be wide in regions where the $b_{12}$ predictions given by the TenMaven are noisy and inaccurate, as demonstrated in the low velocity gradient regions previously highlighted. This ability of being able to predict higher $\sigma_b$ values to indicate noisy, uncertain, and inaccurate predictions of ***b*** is especially helpful and insightful during model deployment. An exception can be found near the hill crest prior to separation in **Figure 8(a)**, where the uncertainty bounds are found to be wide despite close agreement between TenMaven and DNS. This is caused by inaccuracies in the TenMaven predictions of the normal ***b*** components, which led to higher $\sigma_b$ predictions. This approach of applying averaged $\sigma_b$ predictions can raise warning of any inaccurate ***b*** component predictions even if only one component is being analysed. However, it is expected that more accurate uncertainty bounds can be given if separate standard deviations were predicted for each ***b*** component, which will be explored in future work.

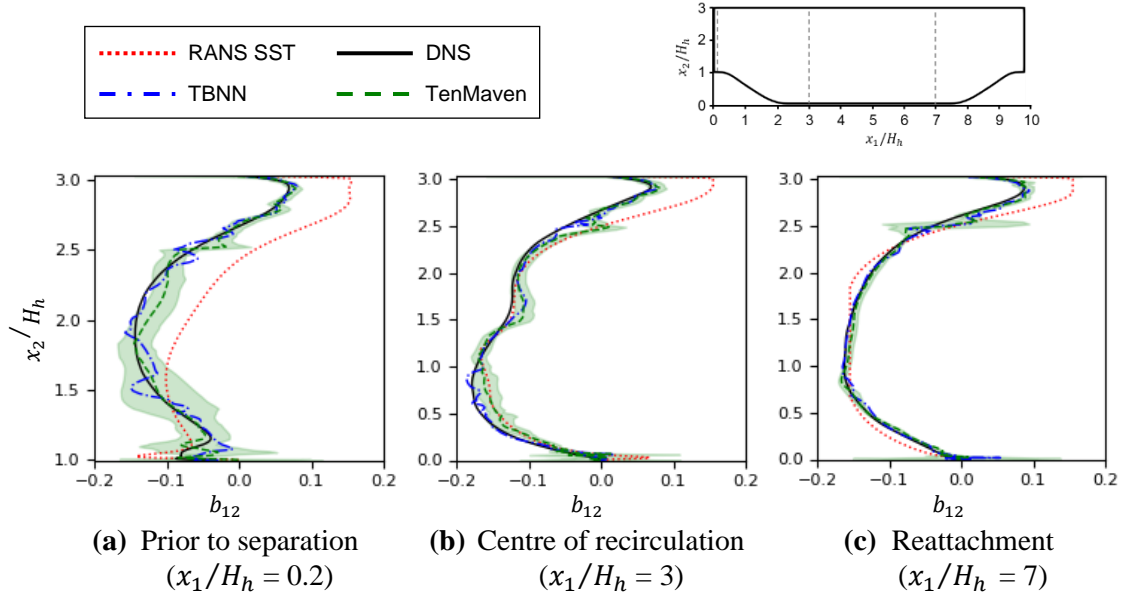

**(a)** Prior to separation $(x_1/H_h = 0.2)$

**(b)** Centre of recirculation $(x_1/H_h = 3)$

**(c)** Reattachment $(x_1/H_h = 7)$

**Figure 8** Line plots of $b_{12}$ predicted by RANS SST, DNS, TBNN, and TenMaven at $x_1/H_h$ locations where **(a)** prior to separation, **(b)** the centre of recirculation, and **(c)** reattachment occurs. These locations are shown as grey dashed lines in the top right schematic. Only results for $b_{12}$ are shown as it is the anisotropy component that governs the behaviour of separated and recirculating flows, which are the features of interest in this test case.

## 4.2 Test Case 2: Flow in a Square Duct

### 4.2.1 Anisotropy Tensor Predictions

Predictions of the anisotropy tensor $\boldsymbol{b}$ given by RANS, DNS, TBNN, and TenMaven for the square duct test case are presented in **Figure 9**. While DNS is shown to predict anisotropic normal $\boldsymbol{b}$ components, RANS fails to do so by predicting zero in the entire domain. This is caused by RANS predicting zero for all normal velocity gradients and its use of BH. As a result, RANS gives isotropic normal Reynolds stresses with Eq. (2), whereas the prediction of accurate secondary flows is dependent on the correct imbalance between normal Reynolds stress components [7]. In contrast, the ML models are able to predict anisotropic normal $\boldsymbol{b}$ components in good agreement with DNS. This is attributed to their use of GEVH to model $\boldsymbol{b}$, which provides greater flexibility compared to the linear dependence on $S$ in BH as discussed previously. Furthermore, these results demonstrate their extrapolation capability, as only periodic hill cases were used for training. The TenMaven is shown to give a minor region of discrepancy near the wall compared to DNS and TBNN, where negative values of $b_{11}$ have been predicted. However, results below show that this error is clearly identified by the $\sigma_{\boldsymbol{b}}$ predictions and would thereby be detected during model deployment.

To quantitatively assess the accuracy of these methods, RMSE for each $\boldsymbol{b}$ component of this test case was also calculated with Eq. (5) and are presented in **Figure 10**. TBNN and TenMaven are respectively shown to have given 68-75% and 55-60% more accurate predictions for the normal components on average compared to RANS. It is believed that the slightly better accuracy given by TBNN compared to the TenMaven is due to exclusive optimisation for $b$ as detailed previously, which may have led to more accurate predictions given by TBNN near the wall.

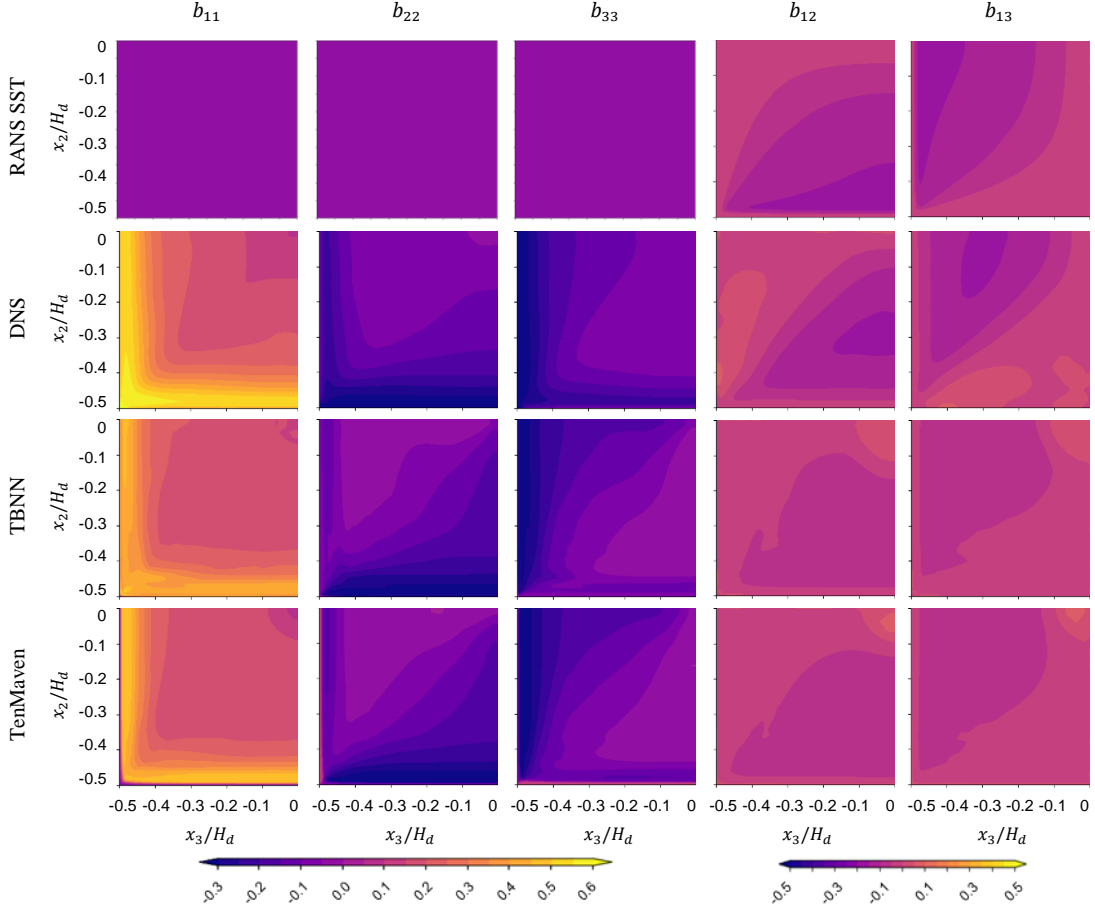

**Figure 9** Anisotropy tensor $b$ component predictions of the square duct case given by RANS SST, DNS, TBNN, and the TenMaven. As the results are symmetric about the wall bisectors where $x_2/H_d = 0$ and $x_3/H_d = 0$, only the lower left quadrants where $x_2/H_d \leq 0$ and $x_3/H_d \leq 0$ are shown. Component $b_{23}$ is omitted as its entire field is zero. Minimum and maximum values of the colour bars are set close to the realizability limits previously stated.

The shear component predictions are also shown in **Figure 9** for completeness. Although the RANS predictions of $b_{12}$ and $b_{13}$ are in satisfactory agreement with the DNS results due to approximate alignment with their corresponding mean strain rate $S$ components, the ML models are also able to predict these more accurately. **Figure 10** shows that the ML models gave 32% and 31% more accurate predictions of $b_{12}$ and $b_{13}$, respectively on average compared to RANS. To assess the overall accuracy of the models for this test case, $RMSE_{all}$ were also calculated with Eq. (6) and are presented in the last column of **Figure 10**. Compared to RANS, the TBNN and TenMaven are shown to predict $b$ significantly more accurately overall by 68% and 55%, respectively.

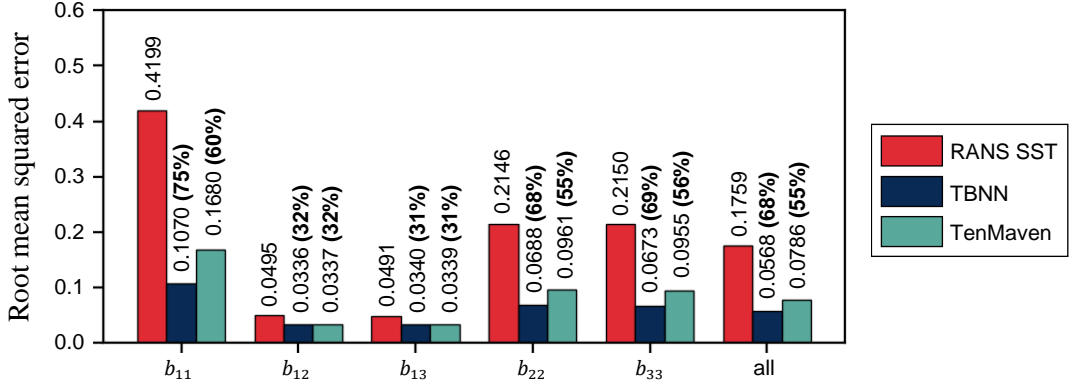

**Figure 10** Root mean squared error of anisotropy tensor $\boldsymbol{b}$ component predictions compared to DNS for the square duct test case. Anisotropy accuracy improvement (%) = $(RMSE_{RANS} - RMSE_{ML}/RMSE_{RANS}) \times 100\%$ given by the ML models are shown in brackets.

#### 4.2.2 TBNN and TenMaven Prediction Similarity

**Figure 11** shows scatterplots of normal anisotropy tensor $\boldsymbol{b}$ predictions given by TBNN and TenMaven plotted against each other. Similar to the periodic hills test case, a positive linear correlation along the identity line can be observed for all components with some points deviating from it. To assess similarity quantitatively, the CCC for each component was calculated and are presented in **Figure 11**. Compared to the periodic hills results, a lower CCC value of 0.61 was calculated for $b_{11}$. This was found to be attributed to the scatter points that correspond to the near-wall predictions given by TenMaven as previously discussed. While this may be considered suboptimal, results below show that the RMSE of $\boldsymbol{b}$ predictions given by TenMaven are still approximately proportional to its $\sigma_{\boldsymbol{b}}$ predictions. High CCC values of 0.86 were calculated for components $b_{22}$ and $b_{33}$.

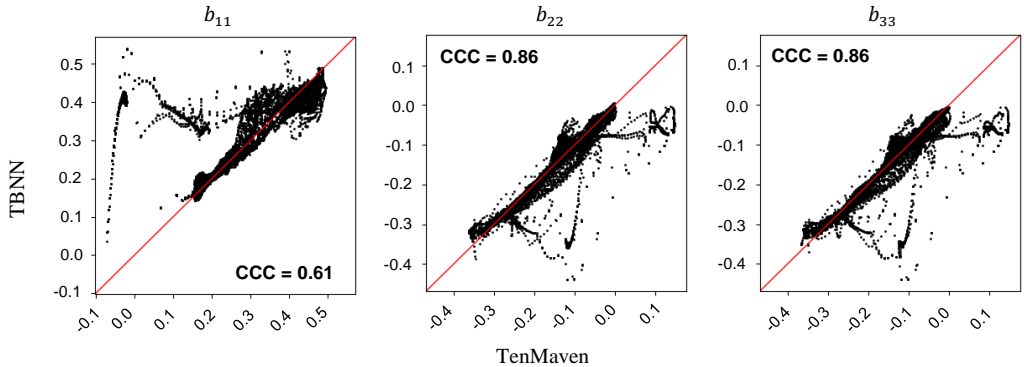

**Figure 11** Scatterplots of anisotropy tensor $\boldsymbol{b}$ components predicted by TBNN (y-axis) vs. TenMaven (x-axis) for the square duct test case. Results for $b_{12}$ and $b_{13}$ are omitted here as the normal components are of greater importance. The red diagonal represents the identity line, where points that lie on it have been predicted identically by both models. The concordance correlation coefficient (CCC) between the two sets of predictions for each component are also shown.

#### 4.2.3 Error Quantification

To investigate the EQ capabilities of $\sigma_{\boldsymbol{b}}$ predictions for this test case, $RMSE^{(q)}$ was calculated with Eq. (7) and plotted against $\sigma_{\boldsymbol{b}}$ as shown in **Figure 12(a)**. A clear positive correlation is demonstrated with a high positive PCC value of 0.93, which supports earlier discussions on their theoretical linear proportionality. Despite suboptimal prediction similarity for $b_{11}$ between TBNN and TenMaven, this finding shows that lower levels of similarity between their $\boldsymbol{b}$ results may be sufficient to achieve this proportionality relationship. An equation of $RMSE^{(q)} = 1.07\sigma_{\boldsymbol{b}} + 0.04$ was found for the least squares best fit line shown in red. **Figure 12(b)** shows surface plots of $\sigma_{\boldsymbol{b}}$ and $RMSE^{(q)}$ on the cross-plane of the square duct. The highest errors are found to occur near the corners of the duct and the wall, which are mostly caused by the lower $b_{11}$ values as previously discussed. However, these coincide with the highest $\sigma_{\boldsymbol{b}}$ predictions as expected. Therefore, this test case also demonstrates that using $\sigma_{\boldsymbol{b}}$ as a proxy variable for $RMSE^{(q)}$ would successfully inform the practitioner of such $\boldsymbol{b}$ prediction errors in the absence of target data.

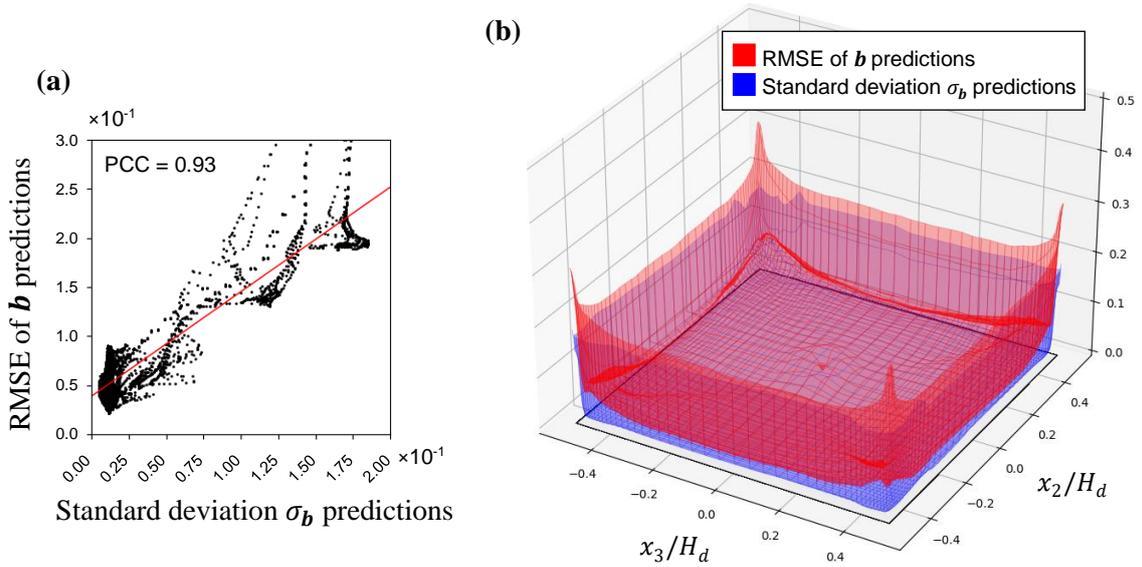

**Figure 12** RMSE of $\boldsymbol{b}$ predictions and standard deviation $\sigma_{\boldsymbol{b}}$ predictions of the square duct test case given by TenMaven plotted **(a)** against each other on a scatter plot, and **(b)** as surfaces on the domain.

### 4.2.4 Uncertainty Quantification

**Figure 13** shows line plots of $b_{11}$ with uncertainty bounds given by $\pm\sigma_{\boldsymbol{b}}$ predicted by the TenMaven at three equally spaced positions along the cross-plane of the square duct. Although the TenMaven profiles are in satisfactory agreement with DNS, the uncertainty bounds mostly do not envelop the DNS profiles in contrast to the previous test case. It is believed that this is caused by the use of a shared standard deviation $\sigma_{\boldsymbol{b}}$ across all $\boldsymbol{b}$ components, where the magnitudes of the other components are much lower as shown in **Figure 9**. Therefore, these findings reiterate the potential merit of predicting separate standard deviations for each $\boldsymbol{b}$ component. Nonetheless, error between the TenMaven and DNS profiles are highest near the wall, which is correlated with $\sigma_{\boldsymbol{b}}$ predictions at all three $x_3/H_d$ positions.

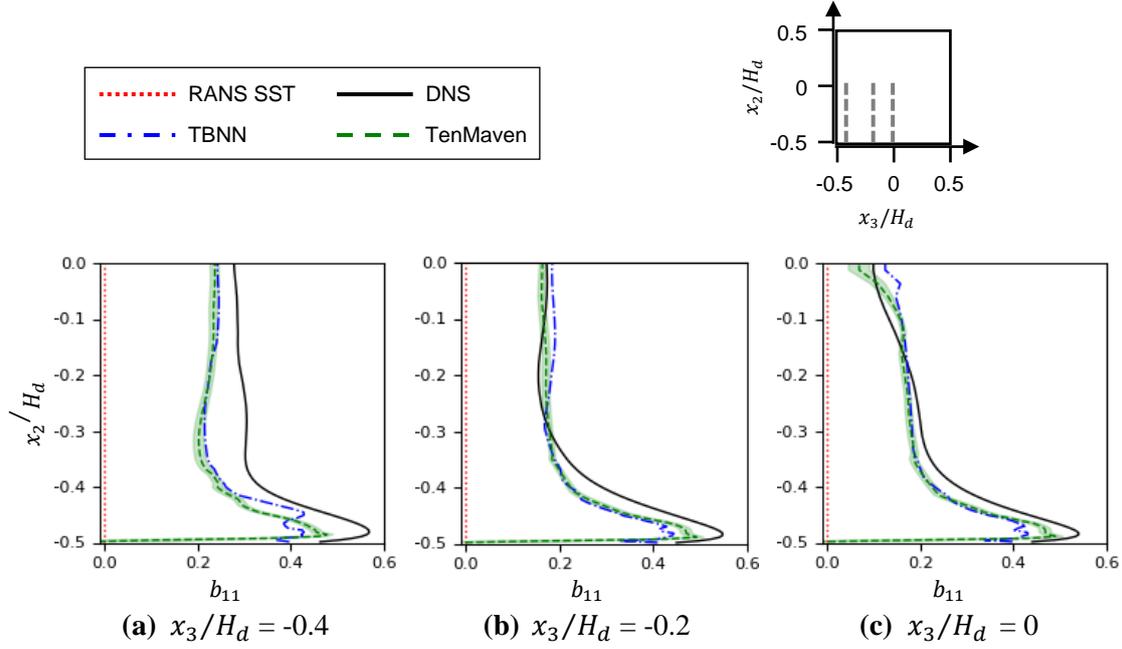

**(a)** $x_3/H_d = -0.4$  **(b)** $x_3/H_d = -0.2$  **(c)** $x_3/H_d = 0$

**Figure 13** Line plots of $b_{11}$ predicted by RANS SST, DNS, TBNN, and TenMaven at $x_3/H_d$ locations along the cross-plane of the square duct. The locations are shown as grey dashed lines in the top right schematic. Only results for $b_{11}$ are presented as the normal anisotropy components govern the behaviour of secondary flows which is the feature of interest in this test case, while components $b_{22}$ and $b_{33}$ have been omitted for brevity. Only the lower left quadrants where $x_2/H_d \leq 0$ and $x_3/H_d \leq 0$ are shown.

## 4.3  Generalisability Comparison

As predicted $\sigma_b$ has been shown to be a valid proxy variable for the RMSE of the TenMaven **b** predictions, an investigation was carried out on whether $\sigma_b$ could identify generalisability error. A comparison of **Figure 5** and **Figure 10** confirms that along with TBNN, the TenMaven also encounters the generalisability problem in its **b** predictions. This is evident in its $RMSE_{all}$ statistic, as the TenMaven gives 10% less accuracy improvement over RANS for **b** predictions of the square duct case compared to the periodic hills case. However, a comparison of $\sigma_b$ surface plots given by TenMaven for the test cases demonstrates that $\sigma_b$ is able to highlight regions of high generalisability error. For the square duct case, **Figure 12(b)** shows that regions of high error near the wall can be easily identified with consistently high values of $\sigma_b$. As generalisability error is caused by the model's failure to respond accurately to flow physics that are different to those in training, $\sigma_b$ and error should both have consistently high levels in the same localised region of flow phenomena where the generalisability problem exists. Although even higher values of $\sigma_b$ are found in the periodic hills result given by **Figure 7(b)**, they are found to be noisy and sparse. Therefore, these results confirm that $\sigma_b$ predictions can highlight generalisability error to make risk-informed decisions in data-driven RANS modelling.

## 5  Conclusion

With the reliability of existing data-driven RANS models in question due to the generalisability problem, an approach based on TBNN with MVENs that embeds uncertainty and error quantification was proposed. Modifications were made to the architecture and loss function of the TBNN to develop the novel MVEN-based RANS closure model, which is referred to as the 'TenMaven'. To assess the TenMaven on cases within and beyond the training data range, a flow over periodic hills case and a flow in a square duct case were respectively chosen for testing. Anisotropy tensor results given by the TenMaven were found to be very similar to those predicted by a TBNN, with high concordance correlation coefficients above 0.85 mostly calculated for the components. This shows that MVENs can preserve the accuracy of the underlying data-driven model. The TenMaven was found to improve the accuracy of anisotropy tensor predictions from RANS by 65% and 55% on average for the respective test cases. While the UQ predictions given by the TenMaven did not envelop much of the DNS profiles for the square duct case, this was successfully achieved for the periodic hills case. It is believed that predicting separate standard deviation values for each anisotropy tensor component could lead to improved UQ predictions. Developing a mixture density network with multiple mixtures may also improve UQ estimates, as predicting multi-modal distributions for the *b* components would be possible.

For EQ purposes, the standard deviation results given by the TenMaven were found to be approximately linearly correlated with the error of its mean anisotropy tensor predictions. This observation is supported by high Pearson correlation coefficient values of 0.88 and 0.93 calculated for the two respective test cases. Hence, this promising result shows that predicted standard deviation from an MVEN is a valid proxy variable for the error of its mean predictions in turbulence modelling. Further exploration of their relationship is necessary for widespread EQ usage however, as slightly different gradients were found for the line of best fit between $\sigma_b$ and pointwise RMSE of the two test cases. Further analysis showed that high, consistent, and smooth values of standard deviation predictions can indicate generalisability error. Therefore, the predicted standard deviation from an MVEN can be used to assess its accuracy, which is expected to be especially insightful in the absence of target data such as during model deployment to give risk-informed predictions.

## 6  Future Directions

Further comprehensive assessments on the capabilities of MVENs will ensure they are accurate and robust to a range of engineering applications. Therefore, applying MVENs to i) other flow cases, ii) alternative data-driven RANS models, and iii) *a posteriori* propagation is necessary to this end. Other modifications may be undertaken to potentially enhance current results, such as performing a hyperparameter study, applying sub-networks, using different MVEN training implementations [75], and training an ensemble [29].

With the reliability of existing data-driven RANS models becoming an important research matter, we bring attention to other UQ methods that have not yet been explored in this field. Such non-Bayesian methods include direct prediction interval construction, and the delta method [64]. Furthermore, some Bayesian approaches have also been rarely applied to this area, such as Monte Carlo dropout which has only been used in one study undertaken by Liu et al. [76] for eddy viscosity modelling. With momentum behind data-driven RANS models, it is expected that they will see increased adoption in industrial applications. Without a universal generalised model, we hope that this study can open further discussions about reliability and works in UQ applied to data-driven RANS modelling.

## Acknowledgements

The authors would like to acknowledge the assistance given by Research IT and the use of the Computational Shared Facility at The University of Manchester.

## Funding

This work was supported by the UK Engineering and Physical Sciences Research Council (grant number EP/W033542/1).

## Declaration of Interests

The authors have no conflicts to disclose.

## References

1. Xiao, H., and Cinnella, P., "Quantification of model uncertainty in RANS simulations: A review", *Progress in Aerospace Sciences* **108**, 1–31 (2019).

2. Durbin, P. A., "Some Recent Developments in Turbulence Closure Modeling", *Annual Review of Fluid Mechanics* **50**, 77–103 (2018).

3. Versteeg, H., and Malalasekera, W., *An Introduction to Computational Fluid Dynamics*, 2nd ed. (Pearson, Essex, UK, 2007).

4. Leschziner, M., *Statistical Turbulence Modelling for Fluid Dynamics – Demystified*, 1st ed. (Imperial College Press, London, UK, 2016).

5. Duraisamy, K., Iaccarino, G., and Xiao, H., "Turbulence Modeling in the Age of Data", *Annual Review of Fluid Mechanics* **51**, 357–377 (2019).

6. Yin, Y., Yang, P., Zhang, Y., Chen, H., and Fu, S., "Feature selection and processing of turbulence modeling based on an artificial neural network", *Physics of Fluids* **32**, 105117 (2020).

7. Wang, J.-X., Wu, J.-L., and Xiao, H., "Physics-informed machine learning approach for reconstructing Reynolds stress modeling discrepancies based on DNS data", *Physical Review Fluids* **2**, 034603 (2017).

8. Berrone, S., and Oberto, D., "An invariances-preserving vector basis neural network for the closure of Reynolds-averaged Navier–Stokes equations by the divergence of the Reynolds stress tensor", *Physics of Fluids* **34**, 095136 (2022).

9. Amarloo, A., Cinnella, P., Iosifidis, A., Forooghi, P., and Abkar, M., "Data-driven Reynolds stress models based on the frozen treatment of Reynolds stress tensor and Reynolds force vector", *Physics of Fluids* **35**, 075154 (2023).

10. Wu, J.-L., Xiao, H., and Paterson, E., "Physics-informed machine learning approach for augmenting turbulence models: A comprehensive framework", *Physical Review Fluids* **3**, 074602 (2018).

11. Chen, Z., and Deng, J., "Data-driven RANS closures for improving mean field calculation of separated flows", *Frontiers in Physics* **12**, 1347657 (2024).

12. Weatheritt, J., and Sandberg, R., "A novel evolutionary algorithm applied to algebraic modifications of the RANS stress–strain relationship", *Journal of Computational Physics* **325**, 22–37 (2016).

13. Schmelzer, M., Dwight, R. P., and Cinnella, P., "Discovery of Algebraic Reynolds-Stress Models Using Sparse Symbolic Regression", *Flow, Turbulence and Combustion* **104**, 579–603 (2020).

14. Ling, J., Kurzawski, A., and Templeton, J., "Reynolds averaged turbulence modelling using deep neural networks with embedded invariance", *Journal of Fluid Mechanics* **807**, 155–166 (2016).

15. McConkey, R., Yee, E., and Lien, F.-S., "Deep structured neural networks for turbulence closure modeling", *Physics of Fluids* **34**, 035110 (2022b).

16. Fang, Y., Zhao, Y., Waschkowski, F., Ooi, A. S. H., and Sandberg, R. D., "Toward More General Turbulence Models via Multicase Computational-Fluid-Dynamics-Driven Training", *AIAA Journal* **61**, 2100–2115 (2023).

17. Goodfellow, I., Bengio, Y., and Courville, A., *Deep Learning*, 1st ed. (The MIT Press, Massachusetts, USA, 2016).

18. Iskhakov, A.S., Tai, C.-K., Bolotnov, I. A., Nguyen, T., Merzari, E., Shaver, D. R., and Dinh, N. T., "Data-Driven RANS Turbulence Closures for Forced Convection Flow in Reactor Downcomer Geometry", *Nuclear Technology* **210**, 1167–1184 (2023).

19. Li, H., Yakovenko, S., Ivashchenko, V., Lukyanov, A., Mullyadzhanov, R., and Tokarev, M., "Data-driven turbulence modeling for fluid flow and heat transfer in peripheral subchannels of a rod bundle", *Physics of Fluids* **36**, 025141 (2024).

20. Ji, Z., and Du, G., "A tensor basis neural network-based turbulence model for transonic axial compressor flows", *Aerospace Science and Technology* **149**, 109155 (2024).

21. Hammond, J., Montomoli, F., Pietropaoli, M., Sandberg, R. D., Michelassi, V., "Machine Learning for the Development of Data-Driven Turbulence Closures in Coolant Systems", *Journal of Turbomachinery* **144**, 081003 (2022).

22. Ellis, C. D., and Xia, H., "Data-driven turbulence anisotropy in film and effusion cooling flows", *Physics of Fluids* **35**, 105114 (2023).

23. Zhao, R., Zhong, S., and You, R., "Application of convolutional neural network for efficient turbulence modeling in urban wind field simulation", *Physics of Fluids* **36**, 105169 (2024).

24. Mandler, H., and Weigand, B., "Generalization Limits of Data-Driven Turbulence Models", *Flow, Turbulence and Combustion* (2024).

25. Abdar, M., Pourpanah, F., Hussain, S., Rezazadegan, D., Liu, L., Ghavamzadeh, M., Fieguth, P., Cao, X. *et al.*, "A review of uncertainty quantification in deep learning: Techniques, applications and challenges", *Information Fusion* **76**, 243–297 (2021).
28

26. Bishop, C. M., *Neural Networks for Pattern Recognition*, 1st ed. (Clarendon Press, Oxford, UK, 1995).

27. Geneva, N., and Zabaras, N., "Quantifying model form uncertainty in Reynolds-averaged turbulence models with Bayesian deep neural networks", *Journal of Computational Physics* **383**, 125–147 (2019).

28. Tang, H., Wang, Y., Wang, T., Tian, L., and Qian, Y., "Data-driven Reynolds-averaged turbulence modeling with generalizable non-linear correction and uncertainty quantification using Bayesian deep learning", *Physics of Fluids* **35**, 055119 (2023).

29. Lakshminarayanan, B., Pritzel, A., and Blundell, C., "Simple and Scalable Predictive Uncertainty Estimation using Deep Ensembles", *Proceedings of the 31st Conference on Neural Information Processing Systems*, Long Beach, USA (2017).

30. Duraisamy, K., "Perspectives on machine learning-augmented Reynolds-averaged and large eddy simulation models of turbulence", *Physical Review Fluids* **6**, 050504 (2021).

31. Agrawal, A., and Koutsourelakis, P.-S., "A probabilistic, data-driven closure model for RANS simulations with aleatoric, model uncertainty", *Journal of Computational Physics* **508**, 112982 (2024).

32. Scillitoe, A., Seshadri, P., and Girolami, M., "Uncertainty quantification for data-driven turbulence modelling with Mondrian forests", *Journal of Computational Physics* **430**, 110116 (2021).

33. Cherroud, S., Merle, X., Cinnella, P., and Gloerfelt, X., "Sparse Bayesian Learning of Explicit Algebraic Reynolds-Stress models for turbulent separated flows", *International Journal of Heat and Fluid Flow* **98**, 109047 (2022).

34. Carlucci, A., Petronio, D., Dellacasagrande, M., Simoni, D., and Satta, F., "Data-Driven Algebraic Models Tuned with a Vast Experimental Database of Separated Flows", *Flow, Turbulence and Combustion* (2024).

35. Gelman, A., Carlin, J.B., Stern, H. S., Dunson, D. B., Vehtari, A., and Rubin, D. B., "Bayesian Data Analysis", 1st ed. (Chapman and Hall/CRC, New York, USA, 1995).

36. Man, A., Jadidi, M., Keshmiri, A., Yin, H., and Mahmoudi, Y., "Optimising a Machine Learning Model for Reynolds Averaged Turbulence Modelling of Internal Flows", *Proceedings of the 16th International Conference on Heat Transfer, Fluid Mechanics and Thermodynamics and Editorial Board of Applied Thermal Engineering*, Online (2022).

37. Zhang, X.-L., Xiao, H., Luo, X., and He, G., "Ensemble Kalman method for learning turbulence models from indirect observation data", *Journal of Fluid Mechanics* **949**, A26 (2022).

38. Gawlikowski, J., Tassi, C. R. N., Ali, M., Lee, J., Humt, M., Feng, J., Kruspe, A., Triebel, R., Jung, P., Roscher, R., Shahzad, M., Yang, W., Bamler, R., and Zhu, X. X., "A survey of uncertainty in deep neural networks", *Artificial Intelligence Review* **56**, S1513–S1589 (2023).

39. Taghizadeh, S., Witherden, F. D., Hassan, Y. A., and Girimaji, S. S., "Turbulence closure modeling with data-driven techniques: Investigation of generalizable deep neural networks", *Physics of Fluids* **33**, 115132 (2021).

40. Maulik, R., Fukami, K., Ramachandra, N., Fukagata, K., and Taira, K., "Probabilistic neural networks for fluid flow surrogate modeling and data recovery", *Physical Review Fluids* **5**, 104401 (2020).

41. Shin, J., Ge, Y., Lampmann, A., and Pfitzner, M., "A data-driven subgrid scale model in Large Eddy Simulation of turbulent premixed combustion", *Combustion and Flame* **231**, 111486 (2021).

42. Nix, D. A., and Weigend, A. S., "Estimating the mean and variance of the target probability distribution", *Proceedings of 1994 IEEE International Conference on Neural Networks*, Orlando, USA (1994).

43. Wilcox, D., *Turbulence Modeling for CFD*, 3rd ed. (s.l.:DCW Industries, Inc, California, USA, 2006).

44. Pope, S. B., *Turbulent Flows*, 1st ed. (Cambridge University Press, Cambridge, UK, 2000).

45. Milani, P. M., Ling, J., and Eaton, J. K., "On the generality of tensor basis neural networks for turbulent scalar flux modeling", *International Communications in Heat and Mass Transfer* **128**, 105626 (2021).

46. McConkey, R., Yee, E., and Lien, F.-S., "On the Generalizability of Machine-Learning-Assisted Anisotropy Mappings for Predictive Turbulence Modelling", *International Journal of Computational Fluid Dynamics* **36**, 555–577 (2022a).

47. Ho, J., Pepper, N., and Dodwell, T., "Probabilistic machine learning to improve generalisation of data-driven turbulence modelling", *Computers and Fluids* **284**, 106443 (2024).

48. Quiñonero-Candela, J., Sugiyama, M., Schwaighofer, A., and Lawrence, N. D., *Dataset Shift in Machine Learning*, 1st ed. (The MIT Press, Massachusetts, USA, 2009).

49. Srivastava, V., and Duraisamy, K., "Generalizable physics-constrained modeling using learning and inference assisted by feature-space engineering", *Physical Review Fluids* **6**, 124602 (2021).

50. Man, A., Jadidi, M., Keshmiri, A., Yin, H., and Mahmoudi, Y., "Non-unique machine learning mapping in data-driven Reynolds averaged turbulence models", *Physics of Fluids* **36**, 095119 (2024).

51. Man, A., Jadidi, M., Keshmiri, A., Yin, H., and Mahmoudi, Y., "A divide-and-conquer machine learning approach for modeling turbulent flows", *Physics of Fluids* **35**, 055110 (2023).

52. Matai, R., and Durbin, P. A., "Zonal Eddy Viscosity Models Based on Machine Learning", *Flow, Turbulence and Combustion* **103**, 93–109 (2019).

53. Oulghelou, M., Cherroud, S., Merle, X., and Cinnella, P., "Machine-learning-assisted Blending of Data-Driven Turbulence Models", *arXiv*:2410.14431 (2024).

54. Cherroud, S., Merle, X., Cinnella, P., and Gloerfelt, X., "Space-dependent Aggregation of Stochastic Data-driven Turbulence Models", *Journal of Computational Physics* **527**, 113793 (2025).

55. Khosravi, A., Nahavandi, S., Creighton, D., and Atiya, A. F., "Comprehensive Review of Neural Network Based-Prediction Intervals and New Advances", *IEEE Transactions on Neural Networks* **22**, No. 9, 1341–1356 (2011).

56. Bishop, C. M., "Mixture Density Networks", Technical report NCRG/94/004, Aston University, Birmingham, UK (1994).

57. Mohri, M., Rostamizadeh, A., and Talwalkar, A., *Foundations of Machine Learning*, 2nd ed. (The MIT Press, Massachusetts, USA, 2018).

58. Pope, S. B., "A more general effective-viscosity hypothesis", *Journal of Fluid Mechanics* **72**, 331–340 (1975).

59. Fu, X., Fu, S., Liu, C., Zhang, M., and Hu, Q., "Data-driven approach for modeling Reynolds stress tensor with invariance preservation", *Computers and Fluids* **274**, 106215 (2024).

60. Li J.-P., Tang, D.-G. T, Yi, C., and Yan, C., "Data-augmented turbulence modelling by reconstructing Reynolds stress discrepancies for adverse-pressure-gradient flows", *Physics of Fluids* **34**, 045110 (2022).

61. Grogan, C., Dutta, S., Tano, M., Dhulipala, S. L. N., and Gutowska, I., "Quantifying Model Uncertainty of Neural-Network based Turbulence Closures", *arXiv*:2412.08818 (2024).

62. Kaandorp, M. L. A., and Dwight, R. P., "Data-driven modelling of the Reynolds stress tensor using random forests with invariance", *Computers and Fluids* **202**, 104497 (2020).

63. Jiang, C., Vinuesa, R., Chen, R., Mi, J., Laima, S., and Li, H., "An interpretable framework of data driven turbulence modeling using deep neural networks," *Physics of Fluids* **33**, 055133 (2021).

64. Dipu Kabir, H. M., Khosravi, A., Hosen, M. A., and Nahavandi, S., "Neural Network-Based Uncertainty Quantification: A Survey of Methodologies and Applications", *IEEE Access* **6**, 36218–36234 (2018).

65. Bishop, C. M., *Pattern Recognition and Machine Learning*, 1st ed. (Springer Science+Business Media, LLC, Singapore, 2006).

66. Alpaydin, E., *Introduction to Machine Learning*, 2nd ed. (MIT Press, Massachusetts, USA, 2010).

67. Xiao, H., Wu, J.-L., Laizet, S., and Duan, L., "Flows over periodic hills of parameterized geometries: A dataset for data-driven turbulence modeling from direct simulations", *Computers and Fluids* **200**, 104431 (2020).

68. McConkey, R., Yee, E., and Lien, F.-S., "A curated dataset for data-driven turbulence modelling", *Scientific Data* **8**, 255 (2021).

69. Pinelli, A., Uhlmann, M., Sekimoto, A., and Kawahara, G., "Reynolds number dependence of mean flow structure in square duct turbulence", *Journal of Fluid Mechanics* **644**, 107–122 (2010).

70. Lin, L. I.-K., "A Concordance Correlation Coefficient to Evaluate Reproducibility", *Biometrics* **45**, 255–268 (1989).

71. Altman, D. G., *Practical Statistics for Medical Research*, 1st ed. (Chapman and Hall/CRC, New York, USA, 1990).

72. Larsen, R.J., and Marx, M.L., *An Introduction to Mathematical Statistics and Its Applications*, 5th ed. (Pearson, Boston, USA 2012).

73. Mandler, H., and Weigand, B., "Feature importance in neural networks as a means of interpretation for data-driven turbulence models", *Computers and Fluids* **265**, 105993 (2023).

74. Parmar, B., Peters, E., Jansen, K. E., Doostan, A., and Evans, J. A., "Generalized Non-Linear Eddy Viscosity Models for Data-Assisted Reynolds Stress Closure", *Proceedings of AIAA SciTech 2020 Forum*, Orlando, USA, Paper No. 2020-0351 (2020).

75. Sluijterman, L., Cator, E., and Heskes, T., "Optimal training of Mean Variance Estimation neural networks", *Neurocomputing* **597**, 127929 (2024).

76. Liu, Y., Hu, R., and Balaprakash, P., "Uncertainty Quantification of Deep Neural Network-Based Turbulence Model for Reactor Transient Analysis", *Proceedings of the ASME 2021 Verification and Validation Symposium*, Online (2021b).

77. Geman, S., Bienenstock, E., and Doursat, R., "Neural Networks and the Bias/Variance Dilemma", *Neural Computation* **4**, 1–58 (1992).

78. He, X., Tan, J., Rigas, G., and Vahdati, M., "On the explainability of machine-learning-assisted turbulence modeling for transonic flows", *International Journal of Heat and Fluid Flow* **97**, 109038 (2022).

79. Ling, J., and Templeton, J., "Evaluation of machine learning algorithms for prediction of regions of high Reynolds averaged Navier Stokes uncertainty", *Physics of Fluids* **27**, 085103 (2015).

# Appendix A: Proportionality between RMSE and Standard Deviation

## A.1 MSE at the Infinite Training Data Limit

Let a multilayer perceptron (MLP) be trained to approximate a function that takes a set of input variables $\boldsymbol{x} \equiv \{x_1, \ldots, x_d\}$ to predict a set of output variables $\boldsymbol{y} \equiv \{y_1, \ldots, y_c\}$. This is usually achieved by using a finite number of entries in a training dataset $\mathcal{D} \equiv \{\boldsymbol{x}^{(q)}, \boldsymbol{y}^{(q)}\}$, where each entry is denoted by index $q = 1, \ldots, N$. Consider MSE that is typically used for such regression problems:

$$MSE = \frac{1}{Nc} \sum_{q=1}^{N} \sum_{k=1}^{c} \left( \hat{y}_k(\boldsymbol{x}^{(q)}, \boldsymbol{w}) - y_k^{(q)} \right)^2 \tag{A.1}$$

where each output variable is denoted by index $k = 1, \ldots, c$. Given an input vector $\boldsymbol{x}^{(q)}$ from a training dataset entry, $\hat{y}_k(\boldsymbol{x}^{(q)}, \boldsymbol{w})$ represents the corresponding predictions of the MLP, which is parameterised by an array of weights and biases $\boldsymbol{w}$. When a training dataset containing a finite number of entries is used, the model's complexity must be limited to control the balance between bias and variance. In practice, this is commonly achieved by limiting the number of hidden layers and nodes or by introducing regularisation [77]. However, at the limit of applying infinite training data (ITD) where $N \to \infty$, both bias and variance can be reduced to zero, as the model can be more flexible by being parameterised with many parameters without overfitting. This limit can be calculated by replacing the finite sum over all training dataset entries in Eq. (A.1) with a double integral over the joint probability density $p(y_k, \boldsymbol{x})$:

$$MSE_\infty = \frac{1}{c} \int \int \sum_{k=1}^{c} (\hat{y}_k(\boldsymbol{x}, \boldsymbol{w}) - y_k)^2 p(y_k, \boldsymbol{x}) \, dy_k d\boldsymbol{x} \tag{A.2}$$

where subscript $\infty$ denotes error at the ITD limit [26]. Hence, the best possible set of $\boldsymbol{w}$ values that can be found during training are those that give the minimum $MSE_\infty$ [57].

## A.2 Minimum RMSE at the Infinite Training Data Limit

The minimum value of $MSE_\infty$ can be obtained if the model gives predictions that satisfy the condition $\delta MSE_\infty / \delta \hat{y}_k = 0$. Let the parameters of such a model be represented as $\boldsymbol{w}^*$. Functionally differentiating Eq. (A.2) with respect to $\hat{y}_k$ and applying the conditional probability rule $p(y_k, \boldsymbol{x}) = p(y_k|\boldsymbol{x})p(\boldsymbol{x})$ gives

$$\frac{\delta MSE_\infty}{\delta \hat{y}_k} = \frac{2}{c} \int (\hat{y}_k(\boldsymbol{x}, \boldsymbol{w}) - y_k) \, p(y_k|\boldsymbol{x}) \, p(\boldsymbol{x}) \, dy_k \tag{A.3}$$

Then setting Eq. (A.3) to 0 and solving for $\hat{y}_k$ results in

$$\hat{y}_k(\boldsymbol{x}, \boldsymbol{w}^*) = \int y_k \, p(y_k|\boldsymbol{x}) \, dy_k \tag{A.4}$$

where $w^*$ represents the set of parameter values that give the minimum $MSE_\infty$ [26]. Eq. (A.4) shows that the minimum $MSE_\infty$ can be obtained if the outputs $\hat{y}_k$ predicted by the model are the conditional average of the target data $y_k$ conditioned on the inputs. Substituting Eq. (A.4) into Eq. (A.2) gives the minimum possible value of $MSE_\infty$ denoted as $MSE_{\infty,min}$:

$$MSE_{\infty,min} = \frac{1}{c} \int \int \sum_{k=1}^{c} (\hat{y}_k(x, w^*) - y_k)^2 p(y_k, x) \, dy_k dx \tag{A.5}$$

By applying the conditional probability rule to Eq. (A.5) and taking the square root, the RMSE calculated using predictions of all possible output values $y_k$ from the model parameterised by $w^*$ for a given input vector $x^{(q)}$ denoted by $RMSE_{|x^{(q)}}$ can be found as

$$RMSE_{|x^{(q)}} = \sqrt{p(x^{(q)})} \sqrt{\frac{1}{c} \int \sum_{k=1}^{c} (\hat{y}_k(x^{(q)}, w^*) - y_k)^2 p(y_k|x^{(q)}) \, dy_k} \tag{A.6}$$

### A.3 MNLL Loss at the Infinite Training Data Limit

The ITD limit of the mean negative log likelihood (MNLL) loss commonly used for mean-variance estimation networks (MVENs) given in Eq. (D.3) can be similarly derived. The first term of Eq. (D.3) can be omitted, as it is an additive constant that is not dependent on the model parameters $w$. By combining the remaining two terms of Eq. (D.3), taking the ITD limit $N \to \infty$, and double integrating over the joint probability density $p(y_k, x)$, the MNLL loss at the ITD limit denoted as $MNLL_\infty$ can be evaluated as

$$MNLL_\infty = \int \int \left( \ln \sigma(x, w) + \frac{1}{2c\sigma(x, w)^2} \sum_{k=1}^{c} (\mu_k(x, w) - y_k)^2 \right) p(y_k, x) \, dy_k dx \tag{A.7}$$

where predictions of the mean $\mu_k$ and standard deviation $\sigma$ are dependent on inputs $x$ and model parameters $w$ of the MVEN [26].

### A.4 Minimum MNLL Loss at the Infinite Training Data Limit

By applying the conditional probability rule and evaluating the functional derivative of $MNLL_\infty$ with respect to $\sigma$, the optimal value of $\sigma$ for a given input vector $x^{(q)}$ can be found. The functional derivative can be evaluated as

$$\frac{\delta MNLL_\infty}{\delta \sigma} = p(x^{(q)}) \int \left( \frac{1}{\sigma(x^{(q)}, w)} - \frac{1}{c\sigma(x^{(q)}, w)^3} \sum_{k=1}^{c} (\mu_k(x^{(q)}, w) - y_k)^2 \right) p(y_k|x^{(q)}) \, dy_k \tag{A.8}$$

Setting Eq. (A.8) to zero, using the integration sum rule, and taking the square root gives

$$\sigma^*_{|\mathbf{x}^{(q)}} = \sqrt{\frac{1}{c} \int \sum_{k=1}^{c} (\mu_k(\mathbf{x}^{(q)}, \mathbf{w}^*) - y_k)^2 \, p(y_k|\mathbf{x}^{(q)}) \, dy_k} \qquad (A.9)$$

where $\mathbf{w}^*$ are the model parameters that predict the optimal standard deviation $\sigma^*_{|\mathbf{x}^{(q)}}$ given $\mathbf{x}^{(q)}$ [26].

**A.5 Proportionality Discussion**

A comparison between Eq. (A.6) and Eq. (A.9) shows that for a given input vector $\mathbf{x}^{(q)}$ and considering all possible output values, the RMSE loss $RMSE_{|\mathbf{x}^{(q)}}$ from an MLP and standard deviation from the MNLL loss $\sigma^*_{|\mathbf{x}^{(q)}}$ of a corresponding MVEN are proportional to each other, provided that the model parameters that give the lowest RMSE and MNLL loss across all possible input $\mathbf{x}$ and output $y_k$ values are used. In practice, obtaining minimum possible RMSE and MNLL loss across the entire input $\mathbf{x}$ and output $y_k$ space is never realised, as the number of training dataset examples is always finite ($N \ll \infty$), and there are always memory limitations on the number of model parameters ($\dim(\mathbf{m}) \ll \infty$) [56]. Hence for a given input vector $\mathbf{x}^{(q)}$, the converged RMSE is an approximation of $RMSE_{|\mathbf{x}^{(q)}}$ from such an ideal MLP, and the converged standard deviation is an approximation of $\sigma^*_{|\mathbf{x}^{(q)}}$ from such an ideal corresponding MVEN. On the basis that $\sigma^*_{|\mathbf{x}^{(q)}}$ is proportional to $RMSE_{|\mathbf{x}^{(q)}}$ from these ideal networks, the predicted standard deviation from an MVEN in practice may be an indicator of the RMSE from a corresponding MLP.

## Appendix B: Input Tensors of the General Effective Viscosity Hypothesis

Based on the Cayley-Hamilton theorem, Pope [58] showed that the input tensors $\mathbf{T}_1, \ldots, \mathbf{T}_{10}$ that form the general effective viscosity hypothesis (GEVH) in Eq. (3) can be written as the following: $\mathbf{T}_1 = \mathbf{S}$, $\mathbf{T}_2 = \mathbf{SR} - \mathbf{RS}$, $\mathbf{T}_3 = \mathbf{S}^2 - (I\{\mathbf{S}^2\}/3)$, $\mathbf{T}_4 = \mathbf{R}^2 - (I\{\mathbf{R}^2\}/3)$, $\mathbf{T}_5 = \mathbf{RS}^2 - \mathbf{S}^2\mathbf{R}$, $\mathbf{T}_6 = \mathbf{R}^2\mathbf{S} + \mathbf{SR}^2 - (2I\{\mathbf{SR}^2\}/3)$, $\mathbf{T}_7 = \mathbf{RSR}^2 - \mathbf{R}^2\mathbf{SR}$, $\mathbf{T}_8 = \mathbf{SRS}^2 - \mathbf{S}^2\mathbf{RS}$, $\mathbf{T}_9 = \mathbf{R}^2\mathbf{S}^2 + \mathbf{S}^2\mathbf{R}^2 - (2I\{\mathbf{S}^2\mathbf{R}^2\}/3)$, $\mathbf{T}_{10} = \mathbf{RS}^2\mathbf{R}^2 - \mathbf{R}^2\mathbf{S}^2\mathbf{R}$, where $\mathbf{S}$ and $\mathbf{R}$ represent the normalised mean strain and rotation rate tensors respectively, $\mathbf{I}$ denotes Kronecker delta, and parentheses $\{\ldots\}$ denote the trace.

## Appendix C: Scalar Inputs

### C.1 Invariant Scalar Inputs

The scalar inputs used in the present work denoted by feature numbers $\lambda_1$ to $\lambda_{47}$ in **Table C.1** include: the first invariants (*i.e.*, traces) of the integrity basis formed by the normalised mean strain rate tensor $\mathbf{S}$, normalised mean rotation rate tensor $\mathbf{R}$, and antisymmetric tensors associated with the pressure gradient $\mathbf{A}_p$ and TKE gradient $\mathbf{A}_k$. The pressure gradient $\nabla \bar{p}$ and TKE gradient $\nabla k$ were normalised and transformed into tensors $\mathbf{A}_p$ and $\mathbf{A}_k$ using the methods detailed in Wu et al. [10].

**Table C.1** Minimal integrity basis tensors formed by symmetric tensor $\mathbf{S}$ and antisymmetric tensors $\mathbf{R}$, $\mathbf{A}_p$, and $\mathbf{A}_k$. The trace of the basis tensors denoted by $\lambda_1$ to $\lambda_{47}$ are used as scalar input features.

An asterisk ($*$) next to a basis tensor indicates that products formed by the cyclic permutation of antisymmetric tensors are also basis tensors (*e.g.*, $\boldsymbol{R}^2\boldsymbol{A_p}\boldsymbol{S}^*$ represents $\boldsymbol{R}^2\boldsymbol{A_p}\boldsymbol{S}$ and $\boldsymbol{A_p^2}\boldsymbol{R}\boldsymbol{S}$).

| Feature number | Basis tensors |
|---|---|
| $\lambda_1$ to $\lambda_2$ | $\boldsymbol{S}^2, \boldsymbol{S}^3$ |
| $\lambda_3$ to $\lambda_5$ | $\boldsymbol{R}^2, \boldsymbol{A_p^2}, \boldsymbol{A_k^2}$ |
| $\lambda_6$ to $\lambda_{14}$ | $\boldsymbol{R}^2\boldsymbol{S}, \boldsymbol{R}^2\boldsymbol{S}^2, \boldsymbol{R}^2\boldsymbol{S}\boldsymbol{R}\boldsymbol{S}^2, \boldsymbol{A_p^2}\boldsymbol{S}, \boldsymbol{A_p^2}\boldsymbol{S}^2,$ $\boldsymbol{A_p^2}\boldsymbol{S}\boldsymbol{A_p}\boldsymbol{S}^2, \boldsymbol{A_k^2}\boldsymbol{S}, \boldsymbol{A_k^2}\boldsymbol{S}^2, \boldsymbol{A_k^2}\boldsymbol{S}\boldsymbol{A_k}\boldsymbol{S}^2$ |
| $\lambda_{15}$ to $\lambda_{17}$ | $\boldsymbol{R}\boldsymbol{A_p}, \boldsymbol{A_p}\boldsymbol{A_k}, \boldsymbol{R}\boldsymbol{A_k}$ |
| $\lambda_{18}$ to $\lambda_{41}$ | $\boldsymbol{R}\boldsymbol{A_p}\boldsymbol{S}, \boldsymbol{R}\boldsymbol{A_p}\boldsymbol{S}^2, \boldsymbol{R}^2\boldsymbol{A_p}\boldsymbol{S}^*, \boldsymbol{R}^2\boldsymbol{A_p}\boldsymbol{S}^{2*}, \boldsymbol{R}^2\boldsymbol{S}\boldsymbol{A_p}\boldsymbol{S}^{2*},$ $\boldsymbol{R}\boldsymbol{A_k}\boldsymbol{S}, \boldsymbol{R}\boldsymbol{A_k}\boldsymbol{S}^2, \boldsymbol{R}^2\boldsymbol{A_k}\boldsymbol{S}^*, \boldsymbol{R}^2\boldsymbol{A_k}\boldsymbol{S}^{2*}, \boldsymbol{R}^2\boldsymbol{S}\boldsymbol{A_k}\boldsymbol{S}^{2*},$ $\boldsymbol{A_p}\boldsymbol{A_k}\boldsymbol{S}, \boldsymbol{A_p}\boldsymbol{A_k}\boldsymbol{S}^2, \boldsymbol{A_p^2}\boldsymbol{A_k}\boldsymbol{S}^*, \boldsymbol{A_p^2}\boldsymbol{A_k}\boldsymbol{S}^{2*}, \boldsymbol{A_p^2}\boldsymbol{S}\boldsymbol{A_k}\boldsymbol{S}^{2*}$ |
| $\lambda_{42}$ | $\boldsymbol{R}\boldsymbol{A_p}\boldsymbol{A_k}$ |
| $\lambda_{43}$ to $\lambda_{47}$ | $\boldsymbol{R}\boldsymbol{A_p}\boldsymbol{A_k}\boldsymbol{S}, \boldsymbol{R}\boldsymbol{A_k}\boldsymbol{A_p}\boldsymbol{S}, \boldsymbol{R}\boldsymbol{A_p}\boldsymbol{A_k}\boldsymbol{S}^2, \boldsymbol{R}\boldsymbol{A_k}\boldsymbol{A_p}\boldsymbol{S}^2, \boldsymbol{R}\boldsymbol{A_p}\boldsymbol{S}\boldsymbol{A_k}\boldsymbol{S}^2$ |

## C.2 Heuristic Scalar Inputs

The scalar invariants were supplemented with eight normalised heuristic scalars denoted by feature numbers $\lambda_{48}$ to $\lambda_{55}$ in **Table C.2**:

**Table C.2** Supplementary normalised heuristic scalars.

| Feature number | Description | Formula | Feature importance studies |
|---|---|---|---|
| $\lambda_{48}$ | Non-dimensional Q-criterion | $\dfrac{\|R\|^2 - \|S\|^2}{\|R\|^2 + \|S\|^2}$ | He et al. [78] |
| $\lambda_{49}$ | Turbulence intensity | $\dfrac{k}{0.5\bar{u}_i\bar{u}_i + k}$ | Wang et al. [7], Li et al. [60], Ji and Du [20] |
| $\lambda_{50}$ | Turbulence Reynolds number | $\min\left(\dfrac{\sqrt{k}d}{50\nu}, 2\right)$ | Wang et al. [7], Yin et al. [6], Li et al. [60], McConkey et al. [46], Mandler and Weigand [73] |
| $\lambda_{51}$ | Pressure gradient along streamline | $\dfrac{\bar{u}_k \frac{\partial \bar{p}}{\partial x_k}}{\sqrt{\frac{\partial \bar{p}}{\partial x_j}\frac{\partial \bar{p}}{\partial x_j}\bar{u}_i\bar{u}_i} + \left|\bar{u}_l \frac{\partial \bar{p}}{\partial x_l}\right|}$ | Li et al. [60] |
| $\lambda_{52}$ | Ratio of turbulent time scale to mean strain time scale | $\dfrac{\|S\|k}{\|S\|k + \varepsilon}$ | Mandler and Weigand [73] |
| $\lambda_{53}$ | Viscosity ratio | $\dfrac{\nu_t}{100\nu + \nu_t}$ | Yin et al. [6], He et al. [78] |
| $\lambda_{54}$ | Ratio of convection to production of TKE | $\dfrac{\bar{u}_i \frac{\partial k}{\partial x_i}}{\left|\overline{u'_j u'_l}S_{jl}\right| + \left|\bar{u}_l \frac{\partial k}{\partial x_l}\right|}$ | Li et al. [60] |
| $\lambda_{55}$ | Ratio of total Reynolds stresses to normal Reynolds stresses | $\dfrac{\left\|\overline{u'_i u'_j}\right\|}{k + \left\|\overline{u'_i u'_j}\right\|}$ | McConkey et al. [46] |

where brackets $\|\cdot\|$ and $|\cdot|$ represent the Frobenius norm and the vector norm, respectively.

**C.3 Scalar Input Normalisation**

Aside from the turbulence Reynolds number $\lambda_{50}$ which is bounded between 0 and 2, the heuristic scalars were bounded between -1 and 1 with the normalisation procedure detailed in Ling and Templeton [79]. Feature bounding is known to improve model robustness and generalisability as dataset shift may thereby be reduced or circumvented [49]. To ensure equal weighting among input features during training, the 47 invariant inputs must also be bounded to a similar range as the heuristic scalars. This was achieved by the formulae chosen for timescales $t_S$ and $t_R$, which were introduced by Wu et al. [10] and follows the normalisation procedure detailed in Ling and Templeton [79].

# Appendix D: Derivation of the Mean Negative Log Likelihood (MNLL) Loss

To derive the MNLL loss $E$, suppose an MVEN parameterised by an array of weights and biases $\boldsymbol{w}$ is trained to approximate a function that takes a set of input variables $\boldsymbol{x} \equiv \{x_1, \ldots, x_d\}$ to predict a set of target variables $\boldsymbol{y} \equiv \{y_1, \ldots, y_c\}$, where the target variables are denoted by subscript $k = 1, \ldots, c$. By assuming that any target variable $y_k$ given input $\boldsymbol{x}$ has uncertainty that is Gaussian distributed, a prediction of the target variable can be assessed by

$$p(y_k|\boldsymbol{x}) = \frac{1}{\sigma(\boldsymbol{x},\boldsymbol{w})\sqrt{2\pi}} \exp\left(-\frac{(\mu_k(\boldsymbol{x},\boldsymbol{w}) - y_k)^2}{2\sigma(\boldsymbol{x},\boldsymbol{w})^2}\right) \tag{D.1}$$

where the prediction has two parts: a mean $\mu_k$, and a standard deviation $\sigma$, which are both dependent on $\boldsymbol{x}$ and $\boldsymbol{w}$. There is no subscript $k$ on $\sigma$ as given a set of values for inputs $\boldsymbol{x}$, Eq. (D.1) supposes that an averaged standard deviation value is applicable across all mean predictions.

In practice, model training is performed by using a finite set of examples that constitute a training dataset $\mathcal{D} \equiv \{\boldsymbol{x}^{(q)}, \boldsymbol{y}^{(q)}\}$, where its entries are denoted by index $q = 1, \ldots, N$. Eq. (D.1) can be extended to calculate the likelihood $\mathcal{L}$ over all target variables $c$ and training dataset entries $N$:

$$\mathcal{L} = \prod_{q=1}^{N} \frac{1}{\sigma^{(q)^c}(2\pi)^{c/2}} \exp\left(-\frac{1}{2\sigma^{(q)^2}} \sum_{k=1}^{c} \left(\mu_k^{(q)} - y_k^{(q)}\right)^2\right) \tag{D.2}$$

The statistical parameters are now dependent on $\boldsymbol{x}^{(q)}$ and $\boldsymbol{w}$ such that the notation can be updated as $\mu_k^{(q)}(\boldsymbol{x}^{(q)}, \boldsymbol{w})$ and $\sigma^{(q)}(\boldsymbol{x}^{(q)}, \boldsymbol{w})$, which have been omitted in Eq. (D.2) for clarity [26]. Although optimised predictions of $\mu_k^{(q)}$ and $\sigma^{(q)}$ can be found by maximising Eq. (D.2), it is more common and intuitive to minimise the mean negative logarithm of Eq. (D.2) averaged over the number of target variables $c$ and entries $N$ to give the MNLL loss $E$:

$$E = -\frac{\ln \mathcal{L}}{Nc} = \frac{1}{2}\ln(2\pi) + \frac{1}{N}\sum_{q=1}^{N} \ln \sigma^{(q)} + \frac{1}{2Nc}\sum_{q=1}^{N} \frac{1}{\sigma^{(q)2}} \sum_{k=1}^{c} \left(\mu_k^{(q)} - y_k^{(q)}\right)^2 \quad \text{(D.3)}$$

Eq. (D.3) has been rewritten to give the MNLL loss used for the TenMaven shown in Eq. (4).